\documentclass[12pt]{article}

\usepackage{jtb}
\usepackage[dvips]{epsfig,graphicx, color}

\usepackage{times}

{
\newenvironment{sciabstract}{%
\begin{quote} \bf}
{\end{quote}}

\newcounter{lastnote}

\author{}

\date{}

\begin{document}


\baselineskip15pt

\begin{center}

{\bf\LARGE An analysis of B cell selection mechanisms\\[1ex]
in germinal centers}\\[1ex]
{\large Michael E. Meyer-Hermann\footnote{To whom
correspondence should be addressed:
Frankfurt Institute for Advanced Studies (FIAS),
Max von Laue Str.~1, 60438 Frankfurt/Main, Germany,
email: M.Meyer-Hermann@f\/ias.uni-frankfurt.de}$^{\hspace*{0.1ex},*,\dagger}$, 
Philip K.~Maini$^{\dagger}$, Dagmar Iber$^\dagger$\\[3mm]
{\normalsize $^{*}$Frankfurt Institute for Advanced Studies (FIAS),}\\
{\normalsize Frankfurt/Main, Germany}\\[1ex]
{\normalsize $^\dagger$Centre for Mathematical Biology, Mathematical Institute,
Oxford, United Kingdom}}
\end{center}


\noindent
{\bf Abbreviations:} Germinal center (GC), centroblast (CB), centrocyte (CC), follicular dendritic cell (FDC), microtubule organizing center (MTOC)

\begin{sciabstract}
{\bf\large Abstract:}\\
Affinity maturation of antibodies during immune responses is achieved by multiple rounds of somatic hypermutation and subsequent preferential selection of those B cells that express B cell receptors with improved binding characteristics for the antigen. The mechanism underlying B cell selection has not yet been defined. By employing an agent-based model, we show that for physiologically reasonable parameter values affinity maturation can neither be driven by competition for binding sites nor antigen --- even in the presence of competing secreted antibodies. Within the tested mechanisms, only clonal competition for T cell help or a refractory time for the interaction of centrocytes with follicular dendritic cells are found to enable affinity maturation while generating the experimentally observed germinal center characteristics and tolerating large variations in the initial antigen density.
\end{sciabstract}

\section{Introduction}

During the course of an immune response, antibodies evolve that bind with increased affinity to an antigen \cite{jerne51,eisen64}. This phenomenon, termed affinity maturation, is based on multiple rounds of somatic hypermutation targeted to the antibody genes and subsequent selection for increased binding affinity \cite{maclennan00}. Both the latter processes are in general confined to germinal centres (GCs) \cite{jacob91b}. Selection acts on B cells, which express as B cell receptor, the rearranged and possibly mutated antibody gene. Recent experiments suggest that B cells are selected in a clonal competition, since low affinity B cells can persist in the GC reaction when competition is reduced \cite{dalporto02,shih02}. Although the key cellular dynamics of the GC reaction are by now well-characterized \cite{liu91,hollowood92,maclennan94} the mechanism enabling clonal competition and thus B cell selection has not yet been defined and is the subject of this paper.

GCs are initially seeded by a small number of proliferating B cells that bind the antigen with at least low affinity \cite{kroese87,jacob91a,liu91}. After a phase of B cell expansion, somatic hypermutation is initiated in so-called centroblasts (CBs) which are characterized by a low expression of surface IgM. Upon differentiation into so-called centrocytes (CCs) the IgM expression increases and CCs are selected according to the quality of antigen binding. CCs are in a state of activated apoptosis \cite{liu97} and in order to survive they need to acquire antigen and present it to antigen-specific T cells that most probably have entered the GCs together with the antigen specific B cells \cite{maclennan00,aydar05}.

Whether antigen trapped on FDCs as opposed to soluble antigen is critical in the selection process is currently controversial \cite{aydar05,haberman03,kosco03}. Selected CCs either differentiate into memory cells or plasma cells or recycle, i.e.~differentiate back into CBs, which enables them to go through multiple rounds of mutation and selection.

Early models of affinity maturation have focused on the selective effect of clonal competition for FDC-presented antigen \cite{siskind69}. This view has been challenged by the recently observed robustness of affinity maturation towards large variation in the amount of initially deposited antigen \cite{vora97,hannum00,manser04}. Given that secreted antibodies appear to be dispensable for affinity maturation \cite{hannum00} this robustness is unlikely to derive from a competition with emerging serum antibodies \cite{vora97,tarlinton00,iber02}.

The alternative model of a clonal competition for binding sites instead of antigen \cite{kesmir03} requires affinity maturation to strictly depend on antigen being encountered membrane-bound, which is the subject of controversy \cite{haberman03}. Also, antigen localization would have to be restricted to a small number of places within the large dendritic network. Despite many electron microscopy studies, such localization restriction has not yet been reported.

Without competition for access to antigen or FDCs, higher affinity B cell clones could be favoured on the level of antigen binding, either because of differential signaling of B cells in response to different affinities of binding \cite{kouskoff98} or because high affinity B cells require fewer encounters with the FDC to establish an immunological synapse and to extract antigen \cite{batista01,meyer-hermann02_jtb}. Here, immunological synapse refers to the supramolecular segregation of proteins in lymphoid cell-cell contact zones \cite{kupfer03}, which enable long-term signalling and effector functions \cite{huppa03}.

Finally, recruitment of T cell help may be competitive. {\it In vitro} assays show that higher affinity B cells are more able to recruit T cell help \cite{batista98}, which may be crucial in the GC environment, where T cells constitute only 5-10\% of the GC cell population \cite{kelsoe96}.

While experiments have addressed the various hypotheses, they have not yet succeeded in ruling out any of the above mechanisms. This is largely due to the experimental difficulties that are associated with the investigation of such complex systems. We therefore employed an extended version of a previously described agent based model for GC reactions \cite{meyer-hermann02_jtb,meyer-hermann05} which allowed us to investigate the impact of each factor separately. Model assumptions are based on experimental data mostly stemming from lymph nodes of mice or rats and the results are robust against physiologically conceivable variations in the parameter values.

We find that both a competition for access to FDCs as well as a competition for 
antigen (even in the presence of antibodies) does not enable affinity maturation to the experimentally observed degree. On the other hand, both a CC refractory time for interaction with FDCs and competition for T cell help can drive affinity maturation --- especially when the acquisition of antigen and survival signals from the FDCs is uncompetitive. Given that both selection mechanisms enable affinity maturation over a wide range of antigen densities, even if antigen is encountered in soluble form, we suggest either one or both of these to be the physiological mechanism of B cell selection. Experiments are suggested to test this unexpected model prediction.

\section{Model}

\subsection{In silico germinal center simulations}
We employ {\it in silico} simulations of the GC reaction to analyse the B cell selection mechanisms. The simulations are performed using a previously described stochastic hybrid agent-based model \cite{meyer-hermann02_jtb,meyer-hermann05} which has been extended to test the different selection mechanisms. The basic idea is to rebuild a GC {\it in silico} and to study how its spatial organization, dynamics and output depend on the assumptions that are made about the comprised cells and their interactions. The GC is represented by a lattice whose nodes represent cells. With the exception of pre-plasma and memory cells the GC cells are confined to the area of the GC as defined by the mantle zone. The cell-lattice is supplemented by a second grid for soluble signals. These satisfy reaction-diffusion-equations which are solved numerically. Given that the analysis of three-dimensional models leads to similar results \cite{meyer-hermann02_dev} we restrict our analysis to two dimensions in order to gain better statistics because of substantially shorter computation times. We include centroblasts, centrocytes, FDCs and output cells such as plasma and memory cells in all our simulations and study the effect of T helper cells on centrocyte selection in an additional set of simulations. The different cell types are encoded in the simulation by defining differently sized objects that are associated with distinct sets of rules which reflect their biological properties. The rules determine the cellular dynamics by defining motility, cell-cell interactions, cell-cycle times, cellular lifetimes and similar properties. A comprehensive list of these properties is given below. Despite the complexity of the GC reaction only a limited set of parameters is required to respect commonly accepted GC properties and to capture available quantitative experimental data. Reasonable estimates can be obtained for most of the parameters (Table 1), and the few exceptions are studied in detail. While most of the experimental data stem from lymph nodes of mice or rats the model applies to all GCs that exhibit similar GC morphology and cell dynamics.
\begin{table}[ht!]
\scriptsize
\begin{center}
\begin{tabular}{|l|c|c|c|c|} \hline
Parameter and reference     & symbol    & value     & type & references\\
\hline
Lattice constant            &$\Delta x$     &$10 \mu m$ & {\rm f\/ixed} &\\
Lattice dimension           &$D$            &$2$            & {\rm f\/ixed} &\\
Radius of reaction volume   &           &$220 \mu m$    & {\rm f\/ixed} &\\
Shape space dimension       &           &$4$        & {\rm f\/ixed} & \cite{perelson79}\\
&&&&\cite{lapedes01}\\
Width of Gaussian aff\/inity weight function                
&           &$2.8$          & {\rm f\/ixed} & \cite{meyer-hermann01}\\
Duration of optimization phase &           &$48 hr$        &{\rm f\/ixed} & 
\cite{jacob93}\\
&&&&\cite{Pascual94}\\
&&&&\cite{meyer-hermann02_jtb}\\
\hline
Number of seeder clones     &           &$3$        & {\rm f\/ixed} & \cite{kroese87}\\
CB velocity      &$v_{\rm CB}$   &$1.5\mu m/min$ & {\rm f\/ixed} & \cite{miller02,wei03}\\
CB persistence time  &$\Delta t_{\rm CB}$    &$2 min$        & {\rm f\/ixed} & \cite{miller02}\\
&&&&\cite{meyer-hermann05pre}\\
Cell cycle time of CB   &           &$9 hr$         & {\rm f\/ixed} & \cite{liu91}\\
Mutation probability of CB &           &$0.5$          & {\rm f\/ixed} & \cite{berek87}\\
&&&&\cite{nossal92}\\
Duration of CB differentiation to CC &$1/r_{\rm diff}$&          & {\rm variable} & \cite{liu91}\\
&&&&\cite{Meyer-Hermann_ImmunolCellBiol}\\
\hline
CC velocity                &$v_{\rm CC}$   &$5\mu m/min$   & {\rm f\/ixed} & \cite{miller02,wei03}\\
CC persistence time      &$\Delta t_{\rm CC}$&$2 min$    & {\rm f\/ixed} & \cite{miller02}\\
&&&&\cite{meyer-hermann05pre}\\
Duration of FDC-CC contact           &           &$2 hr$         & {\rm f\/ixed} & \cite{vaneijk01}\\
CC refractory time &$\Delta t_{\rm delay}$&$6 min$ & {\rm variable} &\\
Duration of differentiation of selected CC      &           &$7 hr$         & {\rm f\/ixed} & \cite{meyer-hermann02_jtb}\\
Probability of recycling for selected CC                &           &$0.8$          & {\rm f\/ixed} & \cite{han95b}\\
&&&&\cite{meyer-hermann01}\\
CC lifetime                &           &$10 hr$        & {\rm f\/ixed} & \cite{liu94}\\
\hline Number of T cells                &$N_{\rm TC}$   &$10$           & {\rm variable} & \cite{kelsoe96}\\
T cell velocity                &$v_{\rm TC}$   &$10.8\mu m/min$& {\rm f\/ixed} & \cite{miller02,wei03}\\
T cell persistence time                &$\Delta t_{\rm TC}$&$2 min$    & {\rm f\/ixed} & \cite{miller02}\\
&&&&\cite{meyer-hermann05pre}\\
Duration of CC-T cell interaction before apoptosis &$\Delta t_{\rm apop}$&$2.1 hr$ & {\rm variable} &\\
Duration of CC-T cell interaction before selection &$\Delta t_{\rm rescue}$&$2.0 hr$& {\rm variable} &\\
\hline Number of FDCs                &$N_{\rm FDC}$  &$10$           & {\rm variable} & \\
Length of FDC dendrites & $d_{\rm FDC}$  &$10\mu m$      & {\rm variable} & \\
Rate of differentiation signal production by FDCs                &$r_{\rm signal}$&      & {\rm variable} & \cite{meyer-hermann02_jtb}\\
Diffusion constant of signal molecules in tissue                &           &$200\mu m^2/hr$& {\rm f\/ixed} & \cite{meyer-hermann02_jtb}\\
Number of antigen portions per FDC site        &$N_{\rm ag}$   &$0$            & {\rm variable} & \\
Antigen threshold for maximum binding probability            &$\Theta_{\rm ag}$&$=N_{\rm ag}$& {\rm variable} &\\
\hline Antibody production by output cells                &           &$0 M/(hr\,cell)$& {\rm variable} & \cite{randall92}\\
Immune complex association rate                &$k_+$          &$10^6/(M s)$   & {\rm f\/ixed} & \cite{batista98}\\
&&&&\cite{fersht98}\\
Immune complex dissociation rate                &$k_-$      &$10^{-3}/s$    & {\rm f\/ixed} & \cite{batista98}\\
&&&&\cite{fersht98}\\
\hline
\end{tabular}
\caption[]{{\small \textbf{Collection of all parameters of the
model with references to the literature on which the values are
based.} The parameters are classif\/ied in the two categories {\it
fixed} and {\rm variable}. {\it Fixed} parameters remain unchanged
throughout all simulations presented in this work. {\it Variable}
parameters have been varied according to the selection process
under consideration. If a value is stated then this is the value
employed in all selection scenarios where this parameter is f\/ixed.
Symbols correspond to the ones used in the text.}}
\label{parameter}
\end{center}
\end{table}

\subsection{Antibody representation}
In order to analyse B cell affinity maturation we need to define the affinity and thus quality of antibodies. Antibodies are represented in a four-dimensional shape space \cite{perelson79}. According to the principle of complementarity the antigen defines the position of the antibody of maximum affinity to it in the shape space. The distance of an arbitrary antibody to this optimal clone is calculated as a 1-norm in the shape space (the minimum number of mutations to reach the optimal clone) and is assumed to be a measure for its affinity to the antigen. The latter is calculated as a Gaussian function with the distance as argument and a width of 2.8 mutations. The power and the width of the affinity function is calculated on the basis of experimental data relating the number of mutations in GC reactions to the increase in affinity (for more details see \cite{meyer-hermann02_jtb,meyer-hermann04}).

\subsection{Cell motility}
In agreement with recent two-photon imaging data collected {\it in vivo} from mice lymph nodes \cite{miller02} cells are assumed to perform an undirected and active movement. Unless stated otherwise, every cell is attributed to a single node. All cell states include a mean speed $v_{\rm cell}$, a polarity $\vec{p}_{\rm cell}$ and a persistence time $\Delta t_{\rm cell}$, where {\it cell} denotes the different cell types. The probability of displacement of a cell to a next neighbour node is determined by $v_{\rm cell}$. The direction of movement is set by $\vec{p}_{\rm cell}$, and the polarity changes with a rate according to $\Delta t_{\rm cell}$. The new polarity is chosen randomly, i.e.~without any memory effect. The values of these motility parameters are taken from experiment \cite{miller02} and are given in the subsequent list of cell types and in Table 1. However, the two-photon experiment does not necessarily apply to GC reactions, such that this has to be considered as an assumption. Note that the actual movement of the cells may be less than expected from $v_{\rm cell}$ when contact inhibition by other cells suppresses the movement.

\subsection{Cell types}
The following list provides a comprehensive description of all cells and their specificities. All parameters without explicitly given value are of importance for the selection process and are discussed within the results section. 

\noindent
{\bf FDC:} The soma of each FDC is represented as a single lattice node. Each FDC is assumed to have four (two per space dimension) dendrites. Every dendrite is attached to the soma and extends to neighbouring nodes. Given that dendrites are flexible the lattice nodes to which dendrites are attributed are not exclusive, i.e.~other cells can move onto these nodes. This is not possible for other cell types or the FDC soma. FDCs are assumed to produce a differentiation signal for centroblasts (CBs) with a rate $r_{\rm signal}$ (see below). In order to reduce computational efforts FDCs are assumed to be immobile. This is a sufficient approximation because the interaction frequency with centrocytes is dominantly determined by the typical distance between FDC binding sites and CCs. This corresponds to the density of the FDC network which is determined by the number of FDCs, $N_{\rm FDC}$, and the length of the dendrites, $d_{\rm FDC}$. A suitable quantity to measure the density of the FDC network is the number of nodes, $X_{\rm FDC}$, from which access to FDC sites is possible in the GC area. We assume access points to be at the position of the dendrites and on all neighbour lattice points. This yields
\begin{equation}
\label{FDCnetwork}
X_{\rm FDC}\;=\;N_{\rm FDC}\,\frac{6D d_{\rm FDC}}{\Delta x}
\quad,
\end{equation}
with $D$ the dimension of the lattice and $\Delta x=10\mu m$ the
resolution of the lattice (chosen as the average B cell diameter). 
The physiological range is $48 \le X_{\rm FDC}\le 144$;
this corresponds to $100$ FDCs per GC \cite{kesmir99} 
($N_{\rm FDC}=4$ in the two-dimensional simulation) with a total length of dendrites of each FDC 
$2Dd_{\rm FDC}=40-120\mu m$. The FDC number in two dimensions is calculated by comparing the FDC density of a spherical GC in three dimensions to the FDC density in a GC slice of thickness $\Delta x=10\mu m$.
 
\noindent
{\bf Centroblasts:} CBs are a subpopulation of B cells in the GC. CBs divide with a cycle time of 6 hours \cite{liu91} (i.e.~9 hours in two dimensions) and, owing to somatic hypermutation, acquire in each cell division non-silent mutations with probability 0.5 \cite{nossal92}. Mutations are modelled by a jump to a next neighbour in the shape space (in an arbitrary direction); wider jumps are excluded. The transcription factors for susceptibility to apoptosis are already up-regulated in CB \cite{klein03}. We assume CBs not to be affected by these before they differentiate to CCs. Equally, CBs are assumed not to interact with FDCs or T cells, thus neglecting possible rescue mechanisms acting directly on CBs. CBs differentiate to CCs at rate $r_{\rm diff}$ once the local concentration of differentiation signal (which is produced by FDCs and diffuses over the lattice) exceeds a threshold value. Given that neither the exact signal nor its threshold concentration are known the signal production rate by FDC is given in units of this threshold. Because of their larger size we expect CBs to move at a lower mean speed 
($v_{\rm CB}=1.5\mu m/min$ and $\Delta t_{\rm CB}=2min$) than measured for naive B cells \cite{miller02}. Assuming that the mean speed scales according to Stokes' friction law, i.e.~with the inverse ratio of the radii of CBs and naive B cells, we obtain a realistic approximation of real motility properties. This assumption might be wrong in view of observed rather motile large B cells \cite{gunzer04}. However, the motility of centroblasts has no important influence for the selection mechanisms which are primarily related to the centrocyte motility.

\noindent
{\bf Centrocytes:} CCs differentiate from CBs and neither divide nor mutate. CCs are in an activated state of apoptosis and they have a lifetime of 10 hours \cite{liu94}, within which they need to receive a rescue signal to avoid death. Dead CCs are rapidly removed from the lattice assuming a fast clearance of apoptotic bodies. The exact nature of the rescue signal depends on the selection mechanism. CCs can be in one of the following states: {\it unselected, in contact with FDC, selected by FDC, in contact with T cell, selected}. Thus the subpopulation of CCs is rather heterogeneous \cite{klein03}.

{\it Unselected} CCs need to find one of the $X_{\rm FDC}$ sites. Once they have access to an FDC site, they try to bind the antigen. The binding probability is proportional to the antibody-antigen affinity. Unsuccessful CCs have to wait for 6 minutes before they can try binding again. This short refractory time ensures independence of the binding process from the time resolution in the simulations. Note that one investigated selection mechanism consists in a prolongation of this refractory time (see results section). Successful CCs switch to the state {\it in contact with FDC} and are rescued from apoptosis. The CC remains bound to the FDC for two hours \cite{lindhout95}. During this time the rescue signals are thought to be provided by the FDC and the CCs turn into the state {\it selected by FDC}. Note that the FDCs are assumed to provide rescue signals and not to negatively select unsuccessful CCs via for example the FasL-Fas pathway \cite{verbeke99}.

All investigated selection mechanisms act on CCs and define how CCs in the state {\it selected by FDC} reach the state {\it selected}. This process is described when the different selection mechanisms are introduced in the results section.

{\it Selected} CCs further differentiate within 7 hours to pre-plasma cells or memory cells (both summarised as output cells in the following), or they, alternatively, recycle back to CBs. No data yet exists to relate the relative receptor quality of the CC to its probability of entering either differentiation path. In fact, available experimental data \cite{han95b} can be reproduced by assuming that this probability does not depend on receptor quality \cite{meyer-hermann01} and as in \cite{meyer-hermann01} we therefore assume that {\it selected} CCs recycle with probability 0.8 and differentiate to output cells with probability 0.2.

CC motility characteristics are assumed to be identical to those of naive B cells 
\cite{miller02} and to be the same in all states, 
i.e.~$v_{\rm CC}=5\mu m/min$ and $\Delta t_{\rm CC}=2min$.

\noindent
{\bf T cells:} T cells are only included in one set of simulations in which they are part of the selection process of CC. As the relative amount of specific and unspecific T cells is not known, only T cells that are specific for the antigen are included, such that all 
$N_{\rm TC}$ T cells will be able to interact with CCs {\it selected by FDC}. Consequently, the number of included T cells may underestimate the total number of T cells present in the GC. The error introduced by this simplification is restricted to spatial effects of T cells, which can be assumed to be small in view of the comparably small total number of T cells in GCs. When T cells interact with CC that have encountered antigen on FDC they develop a polarity towards the CC. If more than one CC {\it selected by FDC} simultaneously interact with a T cell, the T cell is assumed to polarize towards the CC with highest affinity to the antigen. Without CC the polarity of the T cell is random. The T cell motility characteristics are $v_{\rm TC}=10.8\mu m/min$ and $\Delta t_{\rm TC}=2min$ \cite{miller02}. 

\noindent
{\bf Output cells:} Pre-plasma cells and memory B cells are collected in a cell pool denoted by output cells. These cells are assumed to have the same motility characteristics as CCs. In contrast to CCs, output cells can leave the GC environment when reaching the border of the GC. These cells may be thought of as crossing the mantle zone and entering the marginal zone. The total number of
 output cells produced will be taken as a measure of GC success independent of their presence in the GC at the end of the reaction.

\subsection{Diffusion of molecular signals and soluble antibodies}
The model includes diffusing CB differentiation signals as well as soluble
antibodies. The discretised diffusion equation is solved on the lattice 
with Dirichlet boundary conditions and with constant diffusion coefficient. 
Note that the diffusion coefficients 
$D_{\rm diff}=200\mu m^2/hr$ and $D_{\rm ab}=2000\mu m^2/hr$ 
are chosen to be very small in order to respect the high cell 
density in GC that act as obstacles for free diffusion of 
molecules. However, the results are robust against different 
choices of parameter values. Numerically, the Crank-Nicholson 
and the alternating direction implicit (ADI) method are used 
in two and three dimensions, respectively.

In one selection mechanism antigen masking by antibodies 
is considered. Antibodies $b$ are produced by output cells 
at some rate $r_{\rm ab}$, diffuse on the lattice, and 
locally bind to antigen $a$ on FDCs to form immune complexes 
$c$ according to the rate equation 
\begin{equation}
\label{abag} 
\frac{dc}{dt} = k_+ a(t) b(t) - k_- c(t)
\quad.
\end{equation}
While the affinity of the secreted antibodies will increase during the course of the immune response, we simplified the simulation by using binding rates characteristic of high affinity antibodies, e.g. $k_+=10^6/(M s)$
and $k_-=10^{-3}/s$ \cite{batista98,fersht98}. This simplification is appropriate as we show that even under such conditions antibody production at a physiological rate does not sufficiently increase affinity maturation. The total amount of antigen and of antibody is conserved in reaction (2), but these will nonetheless not remain constant since free antigen is consumed by CCs that bind to FDC, and antibodies are produced by output cells. This equation is solved using a simple Euler method at every node.

\subsection{Initial conf\/iguration}
The GC reaction is seeded by a small number of activated B cells that expand rapidly and fill the GC with about $10^4$ B cells before somatic hypermutation starts \cite{mcheyzer-williams93,toellner02} and enables the evolution of higher affinity antibodies. In general a small number of clones dominate the reaction after this expansion. We start the GC simulation after this expansion phase with 1100 CBs (which correspond to the $10^4$ B cells in three dimensions) that stem from three different low affinity clones (binding probability 0.04). The antibodies of these clones all have a distance of 5 mutations to the optimal clone in the shape space; that is, they require a minimum of 5 mutations to gain the optimal affinity. However, most sequences of mutations will not follow such an optimal antibody affinity evolution path. In fact we observe about 9 mutations in most output cells, which is in agreement with experimental data \cite{kueppers93}. The CBs are distributed randomly within the GC. We assume a polarised morphology of the GC in the sense that the FDCs are placed at arbitrary nodes on 2/3 of the total GC volume (\cite{camacho98}, Fig. 1c,d). If T cells are included then they are distributed randomly within the GC. Note, however, that the bias of BC-flow from dark to light zone induces an inhomogeneous TC distribution which are then concentrated in the outer light zone as observed in experiment \cite{hardie93} without additional assumptions.

\subsection{Simulations and analysis of the results}
In order to achieve comparability between the different selection mechanisms, two parameters that affect CB to CC differentiation were varied within physiological limits, while all other parameters are kept constant throughout all simulations (see Table 1), such that the experimentally determined GC kinetics and dark zone duration were reproduced. These two GC characteristics were chosen since both are well established by experiments and variations of these strongly affect the output of the reaction \cite{meyer-hermann02_jtb}.

We assume the differentiation of CB to CC to be initiated by an FDC-derived signal. A signal for CB to CC differentiation has not yet been identified experimentally. However, this mechanism --- unlike other mechanisms --- gives rise to the experimentally observed GC zoning \cite{meyer-hermann05}, and a realistic GC morphology is important for a realistic investigation of selection mechanisms. Beyond the effects stemming from the spatial organisation of the GC, this parameter does not affect B cell selection. The differentiation signal production rate per FDC, $r_{\rm signal}$, widely determines the duration of the dark zone and is adjusted such that the dark zone vanishes between day 8 and 9 
($\sim$ 200 hours --- full line in Figure 1F, and 
in panel C in Figures 2-11 in the Supplementary Information) after immunization \cite{camacho98}.

The second adjusted parameter, the duration of CB-CC differentiation, $\Delta t_{\rm diff}$, governs the total population dynamics and is chosen such that GC kinetic data \cite{liu91,hollowood92} are reproduced with sufficient accuracy (Figure 1D). The deviation $\eta$ from the data is calculated according to
\begin{equation}
\label{deviation} 
\eta \;=\; \sum_{i=1}^N \frac{1}{N-1}
\frac{\left(x_{\rm exp}(t_i) - n(t_i)/n_{\rm max}\right)^2}{x_{\rm
exp}(t_i)^2} \quad,
\end{equation}
where $N$ is the number of experimental values, $t_i$ denotes the time in the reaction at which the value is taken, $n(t_i)$ is the corresponding volume in the simulation, and $n_{\rm max}$ is the peak population averaged over all 50 simulations. We find acceptable GC kinetics when $\eta$ is smaller than 0.3 (Figure 1E and panel B of Figures 2-11 in the Supplementary Information). Note that $\Delta t_{\rm diff}$ must be less than 7 hours since CBs are found in the light zone as CCs within 6-7 hours \cite{liu91}. A minimal value is not known.

The GC reaction is expected to end after 21 days. Given that the simulation was tuned to reproduce the GC kinetics, the cell number within the GC decreases below 100 cells after 300-400 hours for all analysed selection mechanisms (Figure 1F dotted line and panel C in Figures 2-11 in the Supplementary Information). The final cell numbers at day 21 vary for different selection regimes between almost zero and less than 100 (see the same figures, dashed dotted lines).

There are further characteristics of GC reactions to which the simulation was not fitted, but which had to be reproduced by the simulation in order to ensure the model's physiological relevance. Thus the ratio of CCs to CBs at day 12 after immunization needs to be larger than 2 in order to agree with the experimental observation that CCs greatly outnumber CBs during the GC reaction \cite{maclennan90} and the ratio $q^{12}_6$ of produced output cells (i.e.~pre-plasma and memory B cells) at day 12 to day 6 after immunization should be the order of 6 \cite{han95a} (Figure 1C, and panel D of Figures 2-11 in the Supplementary Information). The factor $q^{12}_6$ can be interpreted as a measure for the steepness of the output production.

Unless stated otherwise all simulations remain in agreement with the aforementioned experimental constraints (compare figures in the Supplementary Information where the full data sets are shown). The selection mechanisms that rescue CCs from apoptosis have to rely on some affinity dependent competition between different CCs to achieve affinity maturation. Therefore, the success of a GC reaction is measured by the affinity of the output cells averaged over all cells produced during the reaction (panel B of Figures 1,2,4-6, and panel E of Figures 2-11 in the Supplementary Information). This quantity is an indicator of the entire GC reaction and is not restricted to arbitrary time points of the reaction. The total number of output cells produced is monitored since a stable population will need to be formed to enable robust antibody production. Note that the full datasets are shown only in the case of site competition. For all other scenarios we show only the most important data and refer to the Supplementary Information for the full datasets.

\section{Results}
\subsection{Competition for binding sites on FDCs}
According to the simplest of all suggested models, B cells are selected only in a competition for access to antigen and/or survival factors that are provided by the FDCs; other potentially limiting factors are ignored (Figure 1A in the Supplementary Information). Accordingly, the number of FDC sites $X_{\rm FDC}$ (see Eq. (1)) on which antigen is presented would be the limiting factor for which different clones would compete. As described in the Model section we assume that CC bind antigen held on FDC with a probability proportional to the affinity of the antibody. Given that no further limiting resources are considered in this very simple model, CC that have been selected by FDCs (e.g. CC {\it selected by FDC} in the language of the model, see Model section) do not need any further steps to be positively selected and can follow the differentiation paths of successfully selected CC. 
\begin{figure} [ht!]
\begin{minipage}{15cm}
\begin{center}
\vspace*{-3mm}
\includegraphics[width=7.0cm]{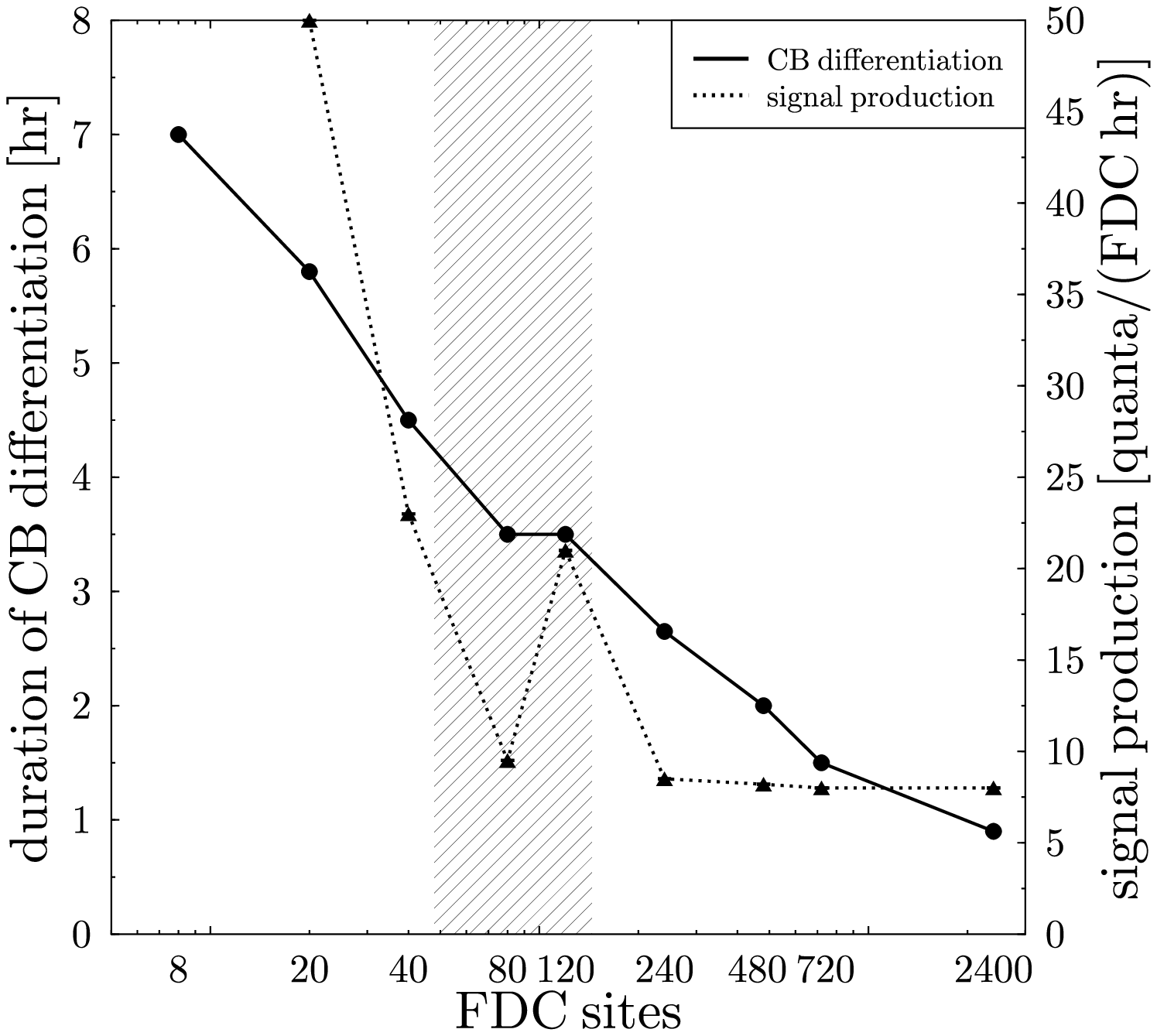}
\includegraphics[width=7.0cm]{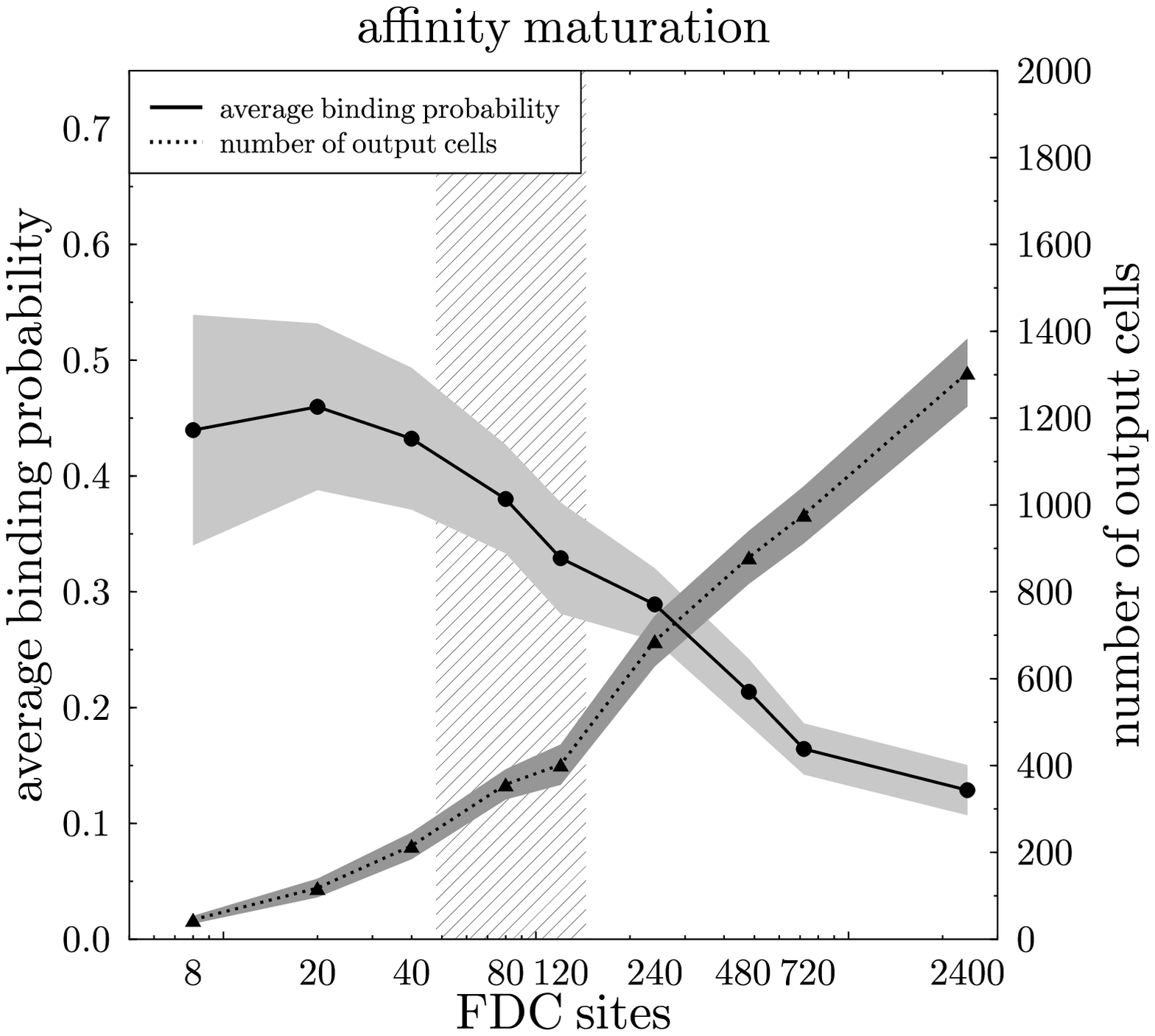}\\[-5mm]
\includegraphics[width=7.0cm]{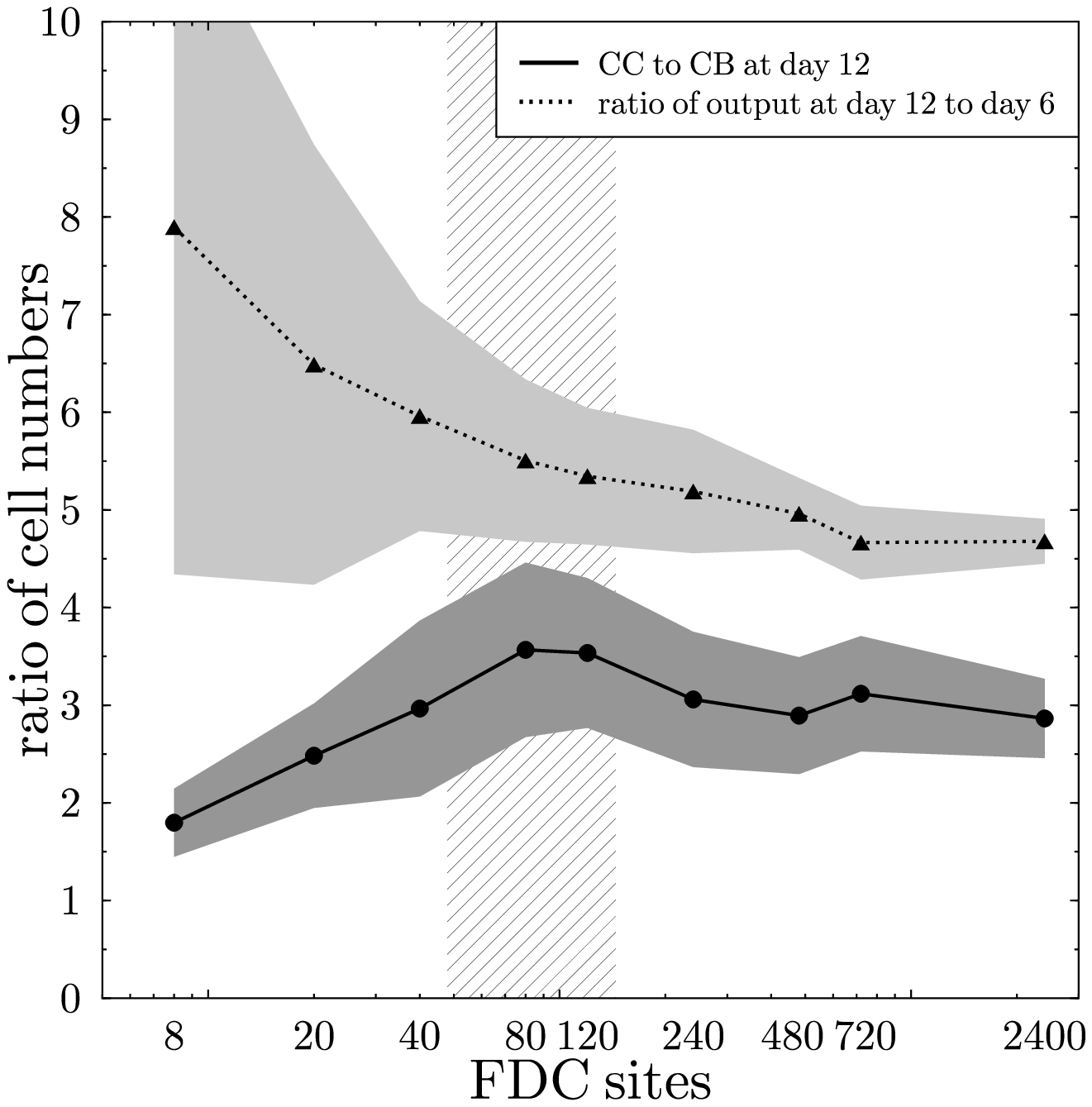}
\includegraphics[width=7.0cm]{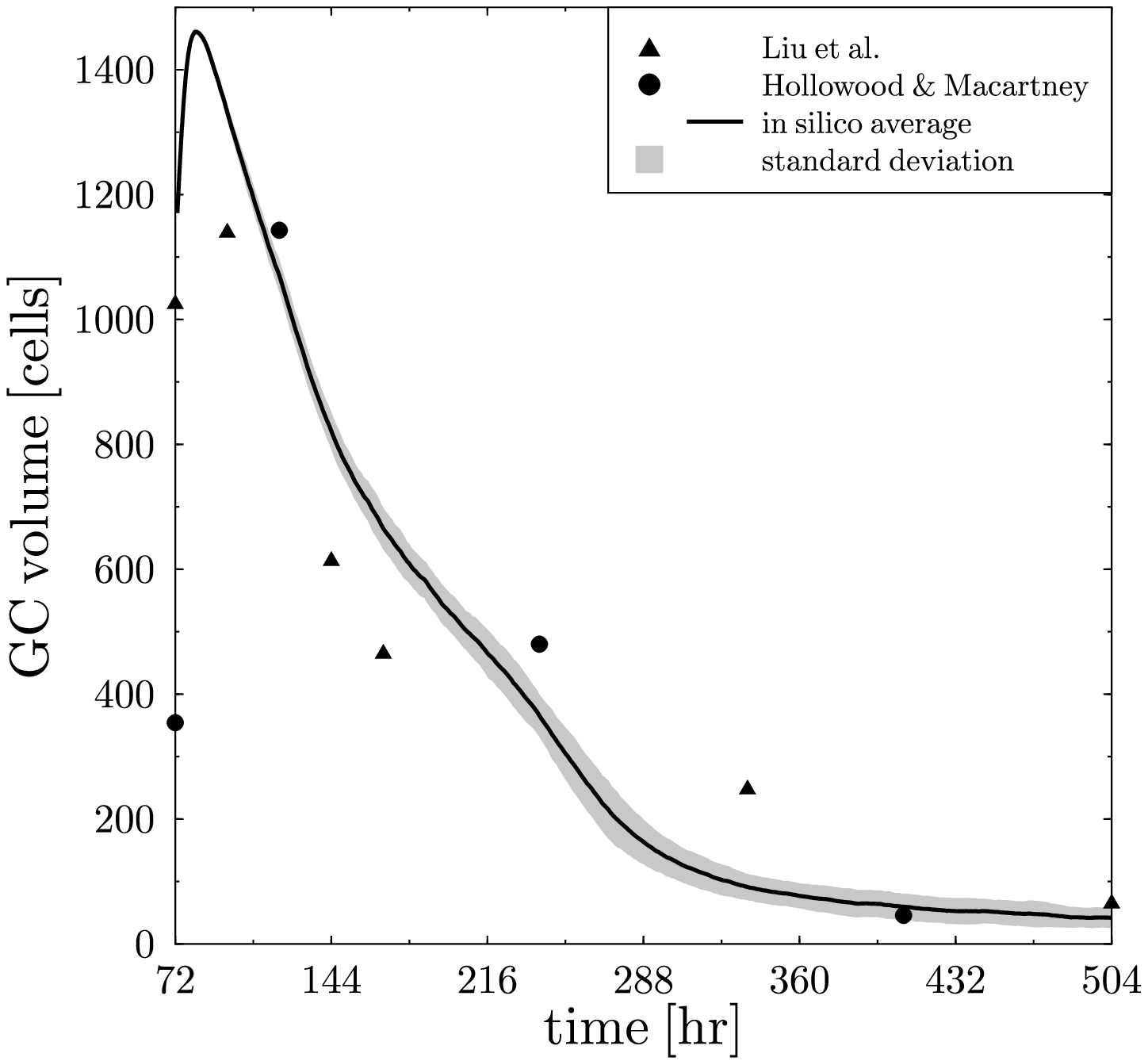}\\[-5mm]
\includegraphics[width=7.0cm]{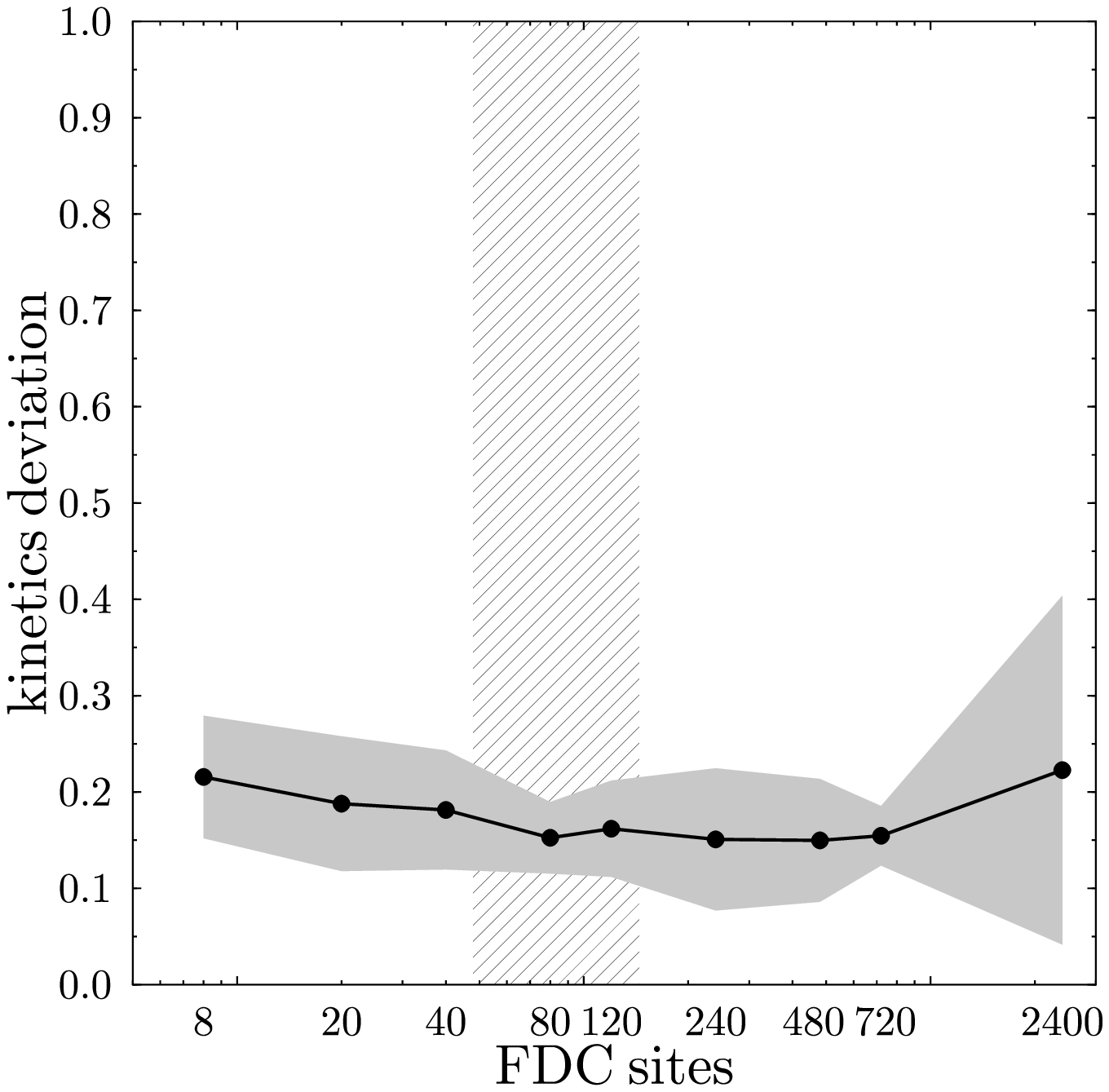}
\includegraphics[width=7.0cm]{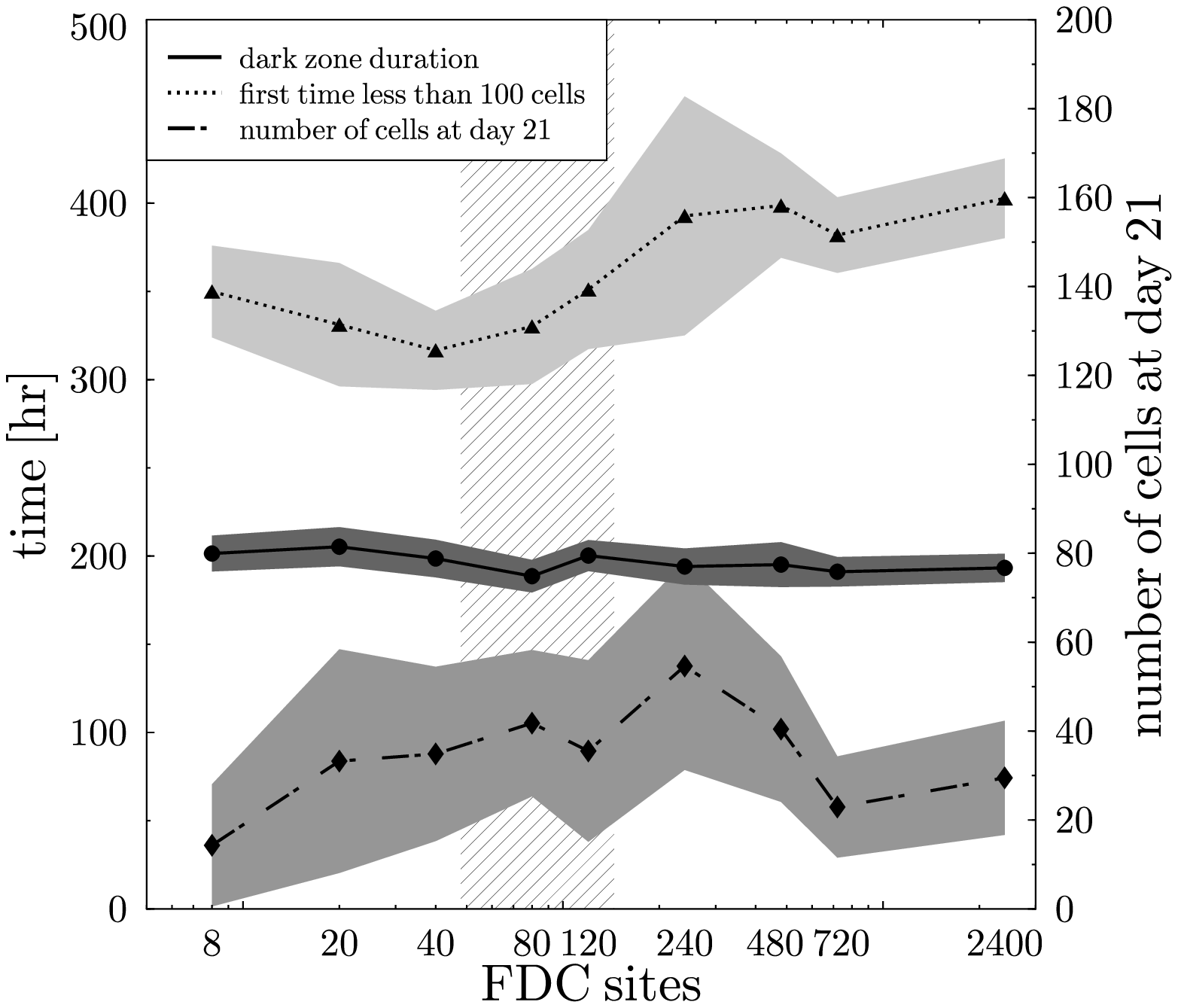}\\[-18.5cm]
\hspace*{-6.7cm}{\bf A}
\hspace*{6.7cm} {\bf B}\\[5.7cm]
\hspace*{-6.8cm} {\bf C}
\hspace*{6.7cm}{\bf D}\\[5.5cm]
\hspace*{-6.8cm} {\bf E}
\hspace*{6.6cm} {\bf F}\\[5.5cm]
\caption[]{\label{Fig_sites} \baselineskip12pt {\small 
\textbf{Competition for FDC sites is too weak for aff\/inity maturation.} 
{\it In silico} GC experiments with different numbers of FDC sites. {\bf (A)} Parameter adaptation: {\it duration of CB differentiation} and differentiation {\it signal production} rate per FDC (note that the jumps in the latter correspond to different numbers of FDCs), {\bf (B)} extent of affinity maturation (measured as average binding probability and number of output cells), {\bf (C)} CC/CB ratio at day 12 of the reaction, ratio $q^{12}_6$ of number of output cells at day 12 to 6, {\bf (D)} GC kinetics taken from 50 simulations with 80 FDC sites including one standard deviation compared to data read off from \cite{liu91,hollowood92}, {\bf (E)} deviation $\eta$ (see Eq. (3)) from experimental GC kinetics, {\bf (F)} end of dark zone, GC population drops below 100 cells for the first time, remaining cell number after 21 days. Vertical bars in {\bf (A,B,C,E,F)} indicate the physiological range. The shaded area denotes one standard deviation in 50 simulations.}}
\end{center}
\end{minipage}
\end{figure}

By variation of the site number $X_{\rm FDC}$ the simulation indeed reveals that affinity maturation is enforced when the number of FDC sites is low (Figure 1B). Affinity maturation can be observed for a physiological number of FDC sites ($48 \le X_{\rm FDC}\le 144$ --- see Eq. (1)); the extent of the process (less than 50\% high affinity output cells) is however rather low compared to other selection mechanisms investigated here and not robust to changes in the number of accessible sites.

The GC characteristics are reproduced {\it in silico}. Thus the CC to CB ratio at day 12, as well as the output ratio at day 12 to day 6, are in agreement with experimental data for the entire physiological range (Figure 1C). The duration of CB-CC differentiation $\Delta t_{\rm diff}$, which decreases from 7 hours to 1 hour with increasing site numbers (Figure 1A), remains within its expected range for the physiological FDC site numbers. Short CB to CC differentiation times had to be employed in order to reproduce the experimentally observed GC kinetics because the selection method is not stringent enough to reduce the volume of the GC reaction considerably.

We conclude that whilst B cell competition for access to FDCs can give rise to the observed GC kinetics and properties, affinity maturation is relatively poor and not robust against small variations in the number of binding sites.

\subsection{Limiting the interactions with FDCs by introducing a refractory time}
If CCs can reinitiate binding to FDCs every 6 minutes (as assumed in our analysis of site competition) then the total number of encounters is considerable during a CC lifetime of 10 hours. Any mechanism that limits the total number of encounters will lower the probability of a successful interaction within a CC� lifetime and will favour higher affinity clones because they have a higher chance of binding successfully in an early attempt. We therefore speculate that an increase in the refractory time ($\Delta t_{\rm delay}$ --- the time during which B cells that failed to bind during a previous contact with a FDC are unable to bind again) will increase the selection pressure (see also Figure 1B in the Supplementary Information). 

Increasing the refractory time from 6 minutes to 4 hours we find that a refractory time as low as 0.5 hours is sufficient to induce affinity maturation of extent similar to that with site competition but for a wide range of FDC site numbers. Thus at $X_{\rm FDC}=120$ (Figure 2), when competition for antigen presenting FDC sites is weak, affinity maturation reaches a level previously only seen at low ($X_{\rm FDC}=40$) site numbers (compare full lines in Figure 1B and 2B). A refractory time of 4 hours leads to rather efficient affinity maturation (60-70\% of all output cells are of high affinity) while allowing an even better reproduction of the other GC characteristics (kinetics, dark zone duration, ratio of CC to CB at day 12 and output ratio at day 12 to day 6 (Figure 2 in the Supplementary Information)) than competition for FDC sites (Figure 1).
\begin{figure} [!ht]
\begin{minipage}{15cm}
\begin{center}
\includegraphics[width=7.0cm]{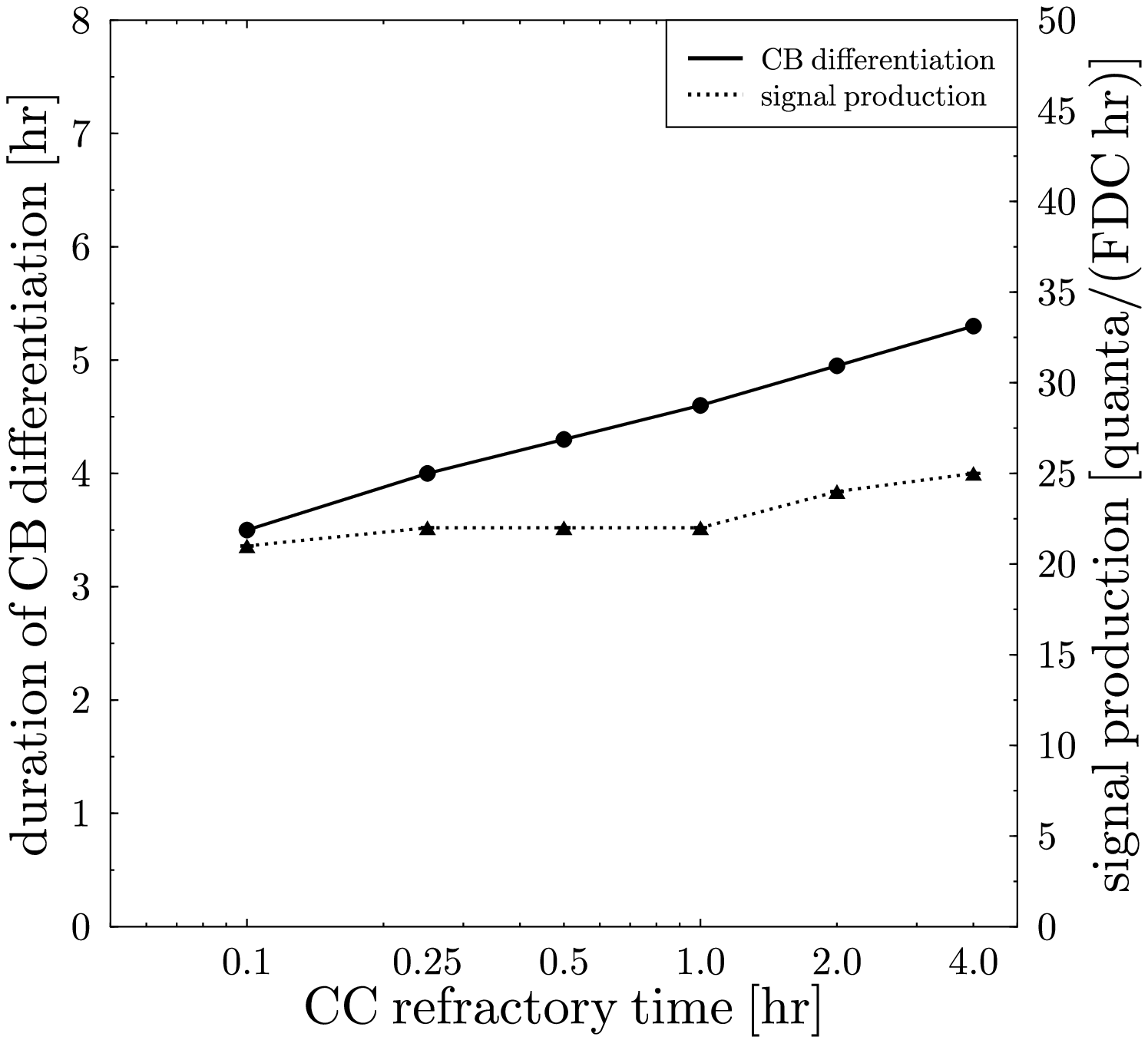}
\includegraphics[width=7.0cm]{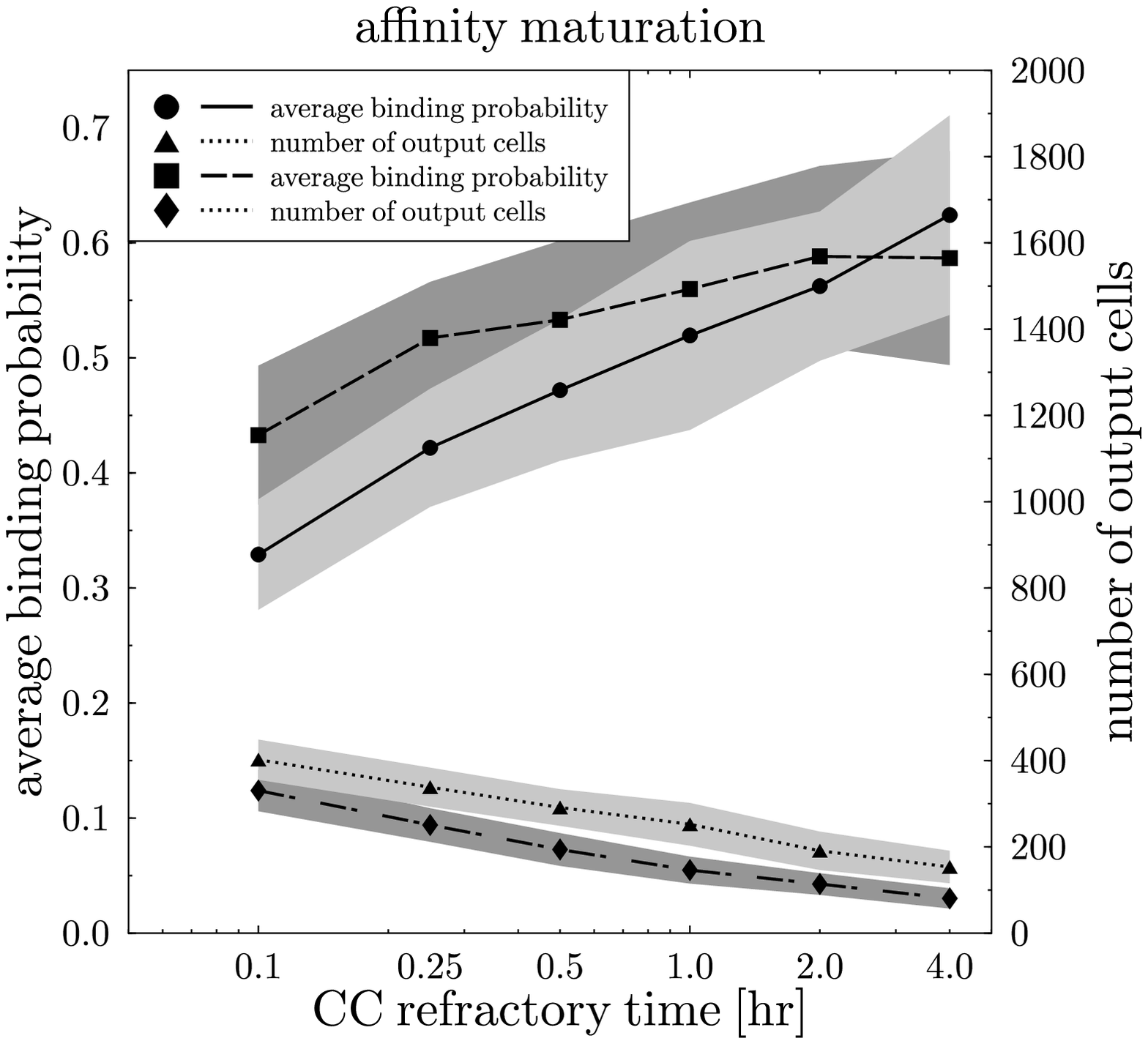}\\[-6.5cm]
\hspace*{-7.0cm}{\bf A}
\hspace*{7.2cm} {\bf B}\\[6.0cm]
\caption[]{\label{Fig_xdelay} \baselineskip12pt {\small
\textbf{A limitation of CC-FDC interaction
        allows for robust aff\/inity maturation.}
{\it In silico}
GC experiments with 10 FDC and 120 FDC sites, and for different refractory times of CC after FDC encounters (see Figure 2 in the Supplementary Information for the full dataset). {\bf (A)} Parameter adaptation: {\it duration of CB differentiation} and differentiation {\it signal production} rate per FDC, {\bf (B)} extent of affinity maturation (measured as average binding probability and number of output cells) with the probability of an optimal clone to bind an antigen presented on FDC of 1 (circles and triangles) and 0.5 (squares and diamonds, and Figure 3 in the Supplementary Information).}}
\end{center}
\end{minipage}
\end{figure}

Given that the binding probability depends on both the affinity dependent binding probability per encounter as well as the number of encounters per CC lifetime, affinity maturation can also be fostered by reducing the binding probability per encounter by a certain factor for all affinities. However, the effect is smaller than in the case of an increased CC refractory time since such a general reduction in the binding probability also strongly reduces the selection probability for high affinity clones. These are little affected by a reduction in the number of encounters since they will bind with high probability on their first or second trial. A combination of both effects can however reduce the refractory time required to gain good affinity maturation to more reasonable physiological values (Figure 2B dashed line, and Figure 3 in the Supplementary Information).

The agreement with general GC properties, the efficient affinity maturation and the robustness to variations in the antigen presenting FDC site number, make this a candidate mechanism for B cell selection {\it in vivo}.

\subsection{Competition for antigen presented on FDCs}
In the two previous scenarios antigen is presented by FDCs on a limited number of sites where it is available in limitless supply. Instead, it has been proposed that antigen is widely spread but that the total amount is limiting such that antigen consumption can drive affinity maturation \cite{sidman75}. Using a sufficiently dense FDC network ($X_{\rm FDC}=120$) to avoid site competition, we can study this selection mechanism {\it in silico} by homogeneously distributing a number of antigen portions, $N_{\rm ag}$, on all FDC sites and by removing one portion from each site on which CCs have successfully interacted with the FDC (see also Figure 1C in the Supplementary Information). The binding probability is expected to decrease with decreasing amounts of antigen remaining on the FDC sites. As a simple approximation we assume a linear decrease of the CC-binding probability for antigen amounts below some threshold $\Theta_{\rm ag}$. For larger antigen amounts the binding probability is constant and solely determined by the antibody-antigen affinity. This assumption corresponds to a saturation of the binding probability when antigen is abundant.

While the selection pressure indeed increases over time as a result of this (Figure 3B), the extent of affinity maturation is rather worse than for site competition alone (compare Figure 1B for $X_{\rm FDC}=120$ and Figure 4B). This is due to the initial abundance of antigen and the inferred low selection pressure which in turn results in the production of a large number of low affinity output cells before antigen consumption can increase the selection pressure (compare Figure 3A and B).
\begin{figure} [!ht]
\begin{minipage}{15cm}
\begin{center}
\includegraphics[width=7.0cm]{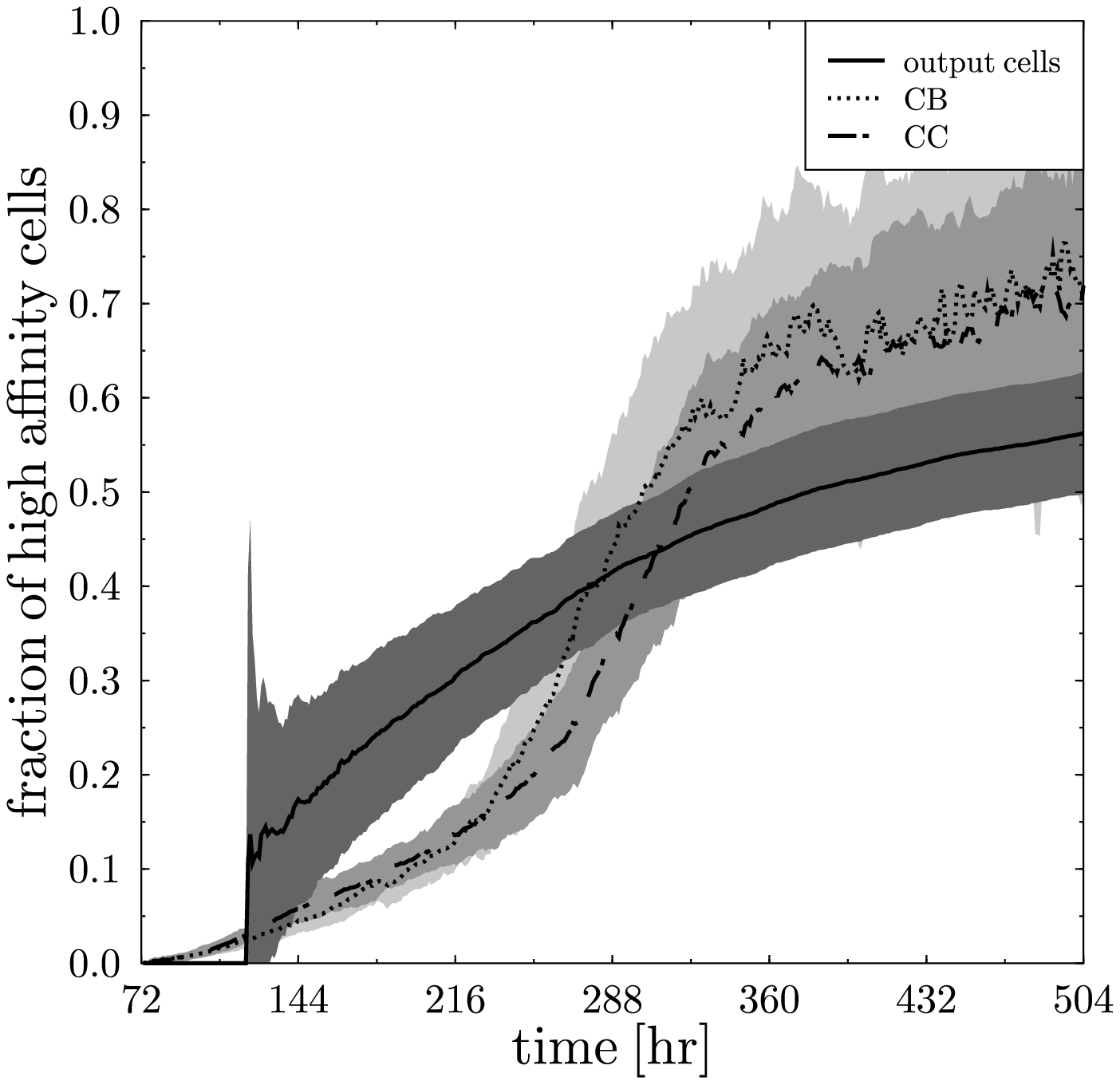}
\includegraphics[width=7.0cm]{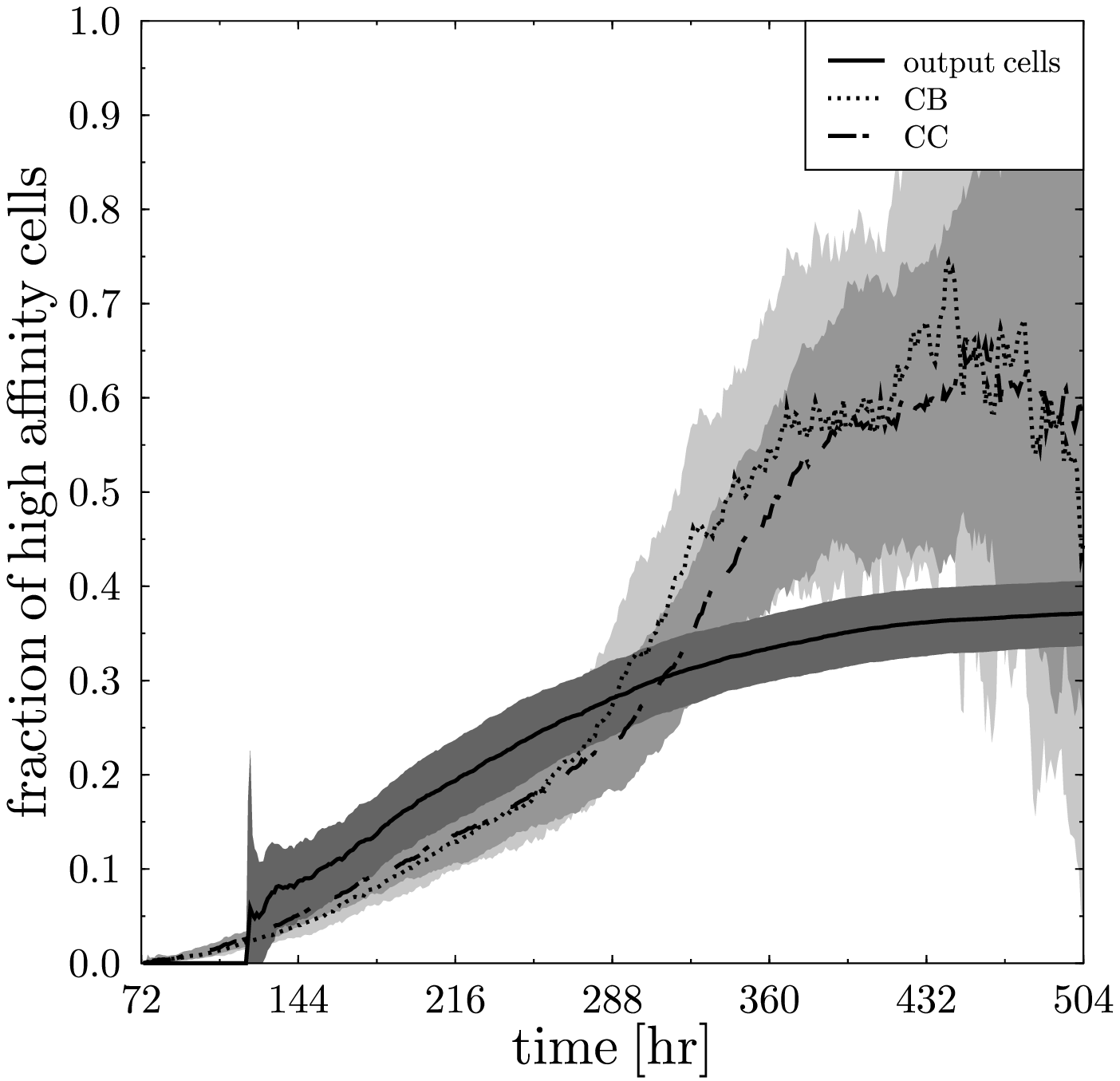}
\includegraphics[width=7.0cm]{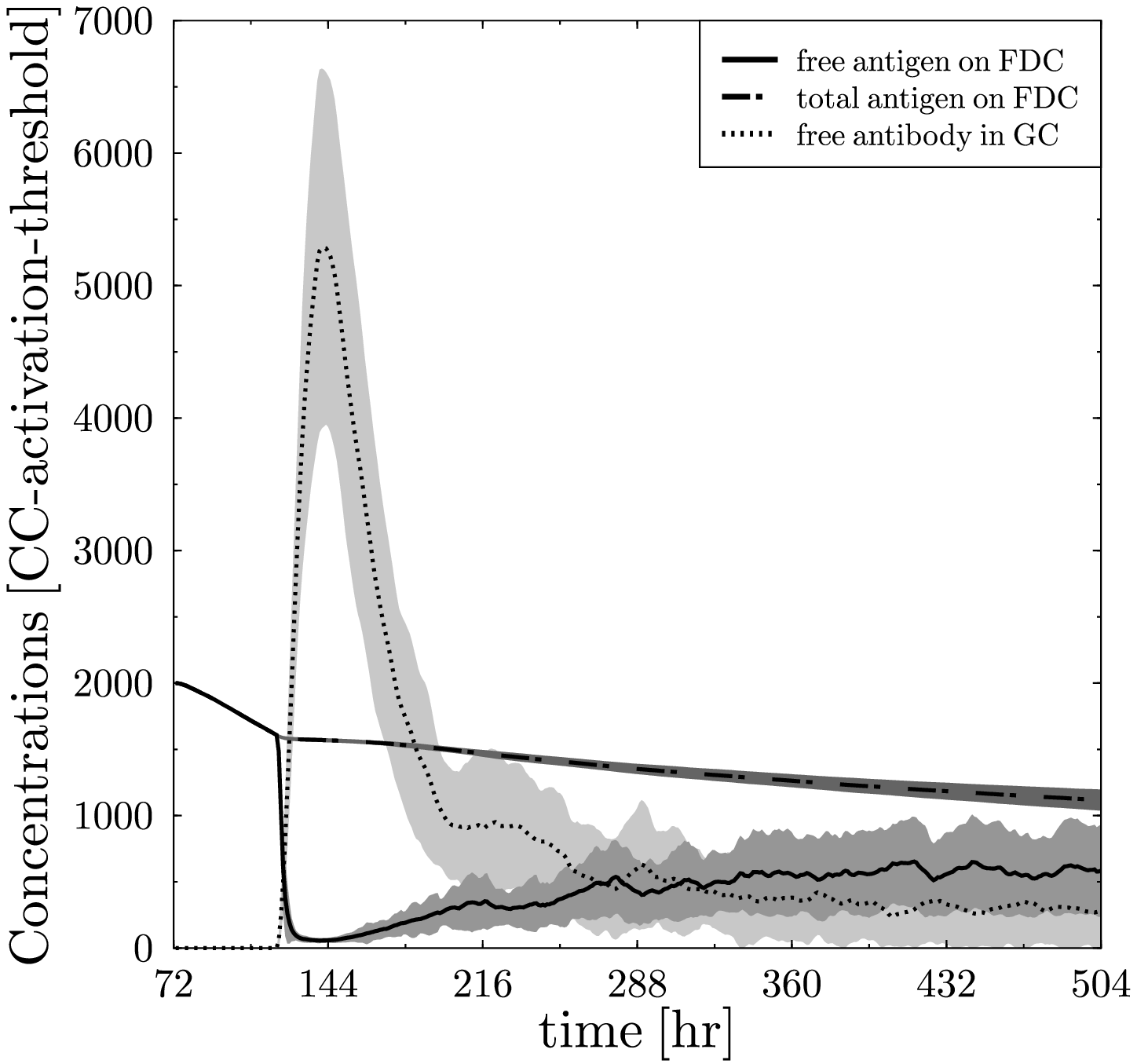}\\[-13.0cm]
\hspace*{-7.0cm}{\bf A}
\hspace*{7.2cm} {\bf B}\\[6.2cm]
\hspace*{-7.6cm} {\bf C}\\[5.8cm]
\caption[]{\label{Fig_affmat_t} \baselineskip12pt {\small 
\textbf{The time course of aff\/inity maturation.} 
The fraction of high affinity cells (defined as cells with antigen binding probability of more than 0.3) is shown for CB, CCs, and the accumulated sum of output cells, separately. In {\bf (A)} selection is based on a refractory time for CC-FDC interaction of 2 hours (taken from Figure 2), while in {\bf (B)} selection is based on competition for antigen with 40 antigen portions per FDC site (taken from Figure 4). In {\bf (C)} the time course of free and total antigen on FDC, and soluble antibody is shown for the same experiment as in {\bf (B)} but including soluble antibody (taken from Figure 5). Soluble antibody induces an early reduction of free antigen and thereby increases competition for antigen as compared to {\bf (B)}. The shaded area denotes one standard deviation in 50 simulations.}}
\end{center}
\end{minipage}
\end{figure}

If the size of the removed antigen portions per CC-FDC interaction is constant, antigen consumption is insufficient to increase the selection pressure in the case of large initial antigen densities (more than 80 antigen portions per FDC site, compare Figures 1B and 4B); at lower antigen density, antigen is rapidly consumed and the GC reaction declines early, before high affinity clones are found (Figure 4, and Figure 4 in the Supplementary Information). In the case of antigen consumption proportional to the level of presented antigen a scenario comparable to antigen masking by antibodies arises (see next section).
\begin{figure} [!ht]
\begin{minipage}{15cm}
\begin{center}
\includegraphics[width=7.0cm]{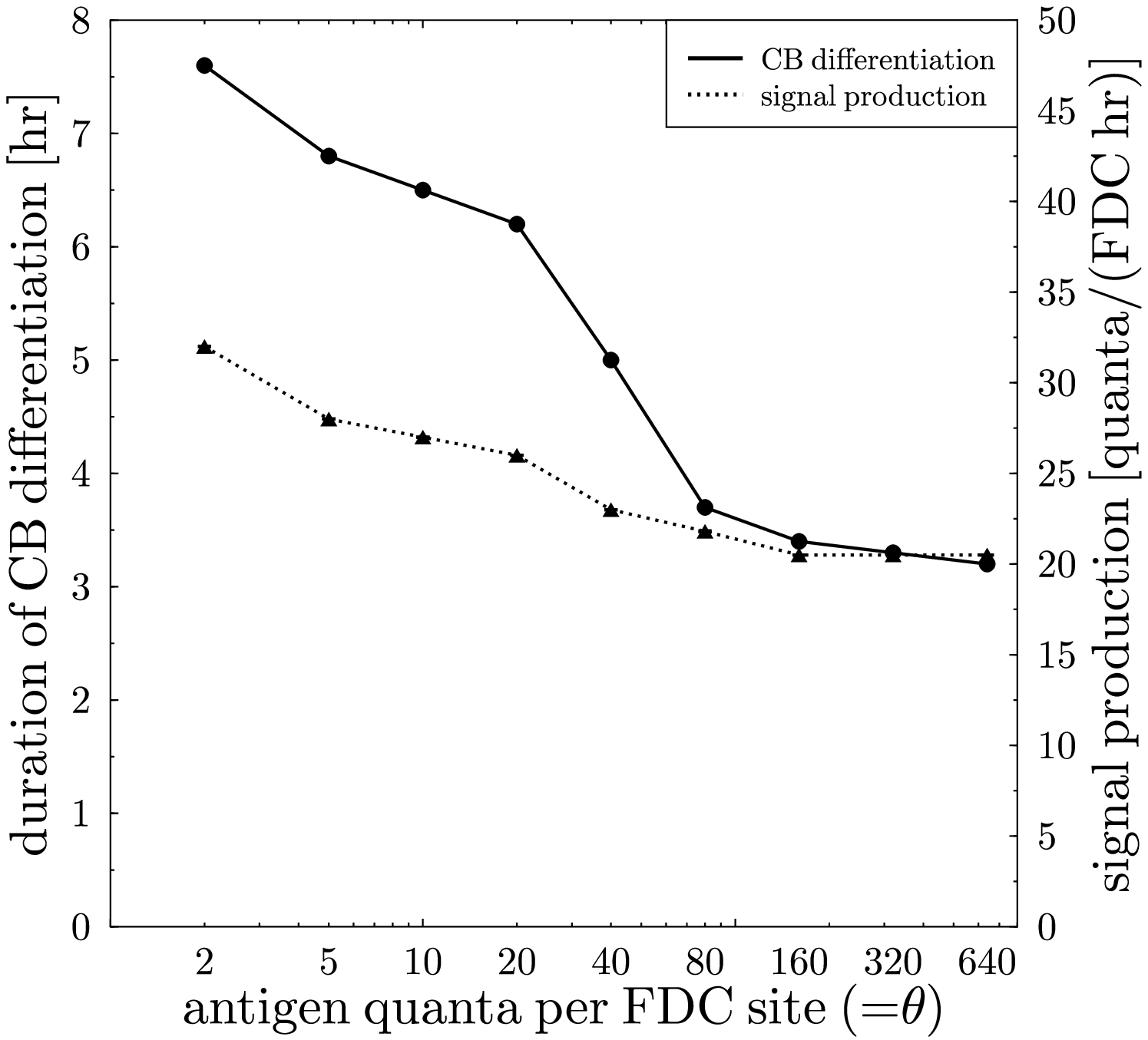}
\includegraphics[width=7.0cm]{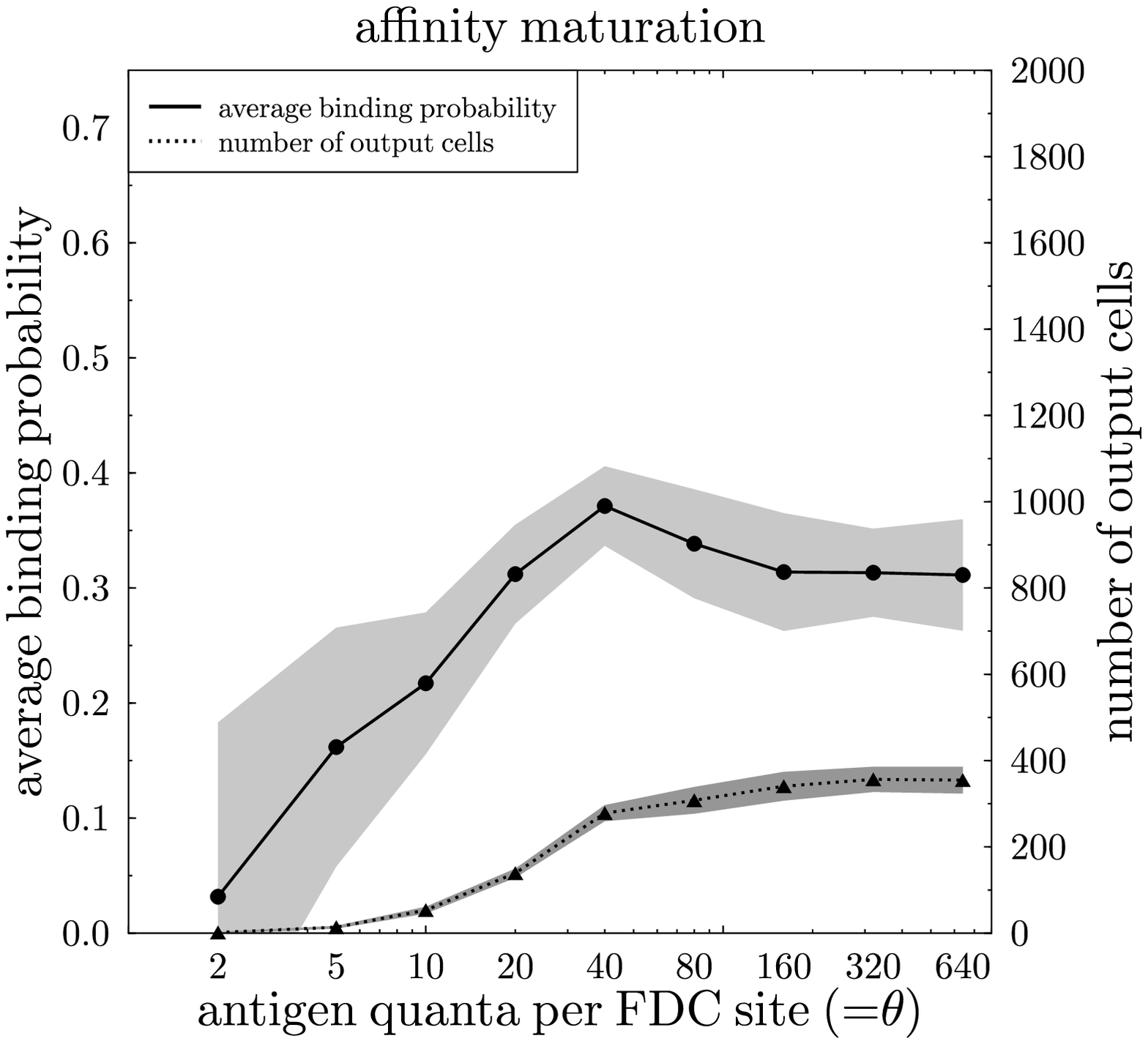}\\[-6.5cm]
\hspace*{-6.5cm}{\bf A}
\hspace*{7.0cm} {\bf B}\\[6.0cm]
\caption[]{\label{Fig_agxtx} \baselineskip12pt {\small
\textbf{Aff\/inity maturation is not induced by competiton for antigen.}
{\it In silico} GC experiments with 120 FDC sites. The antigen binding probability of CC is linearly reduced with the amount of remaining antigen on FDCs (see Figure 4 in the Supplementary Information for the full dataset). {\bf (A)} Parameter adaptation: {\it duration of CB differentiation} and differentiation {\it signal production} rate per FDC, {\bf (B)} extent of affinity maturation (measured as average binding probability and number of output cells).}}
\end{center}
\end{minipage}
\end{figure}

Experiments have revealed a high robustness to variations in the initial antigen density \cite{vora97,hannum00}. If, in the simulation the antigen-threshold to the maximum binding probability $\Theta_{\rm ag}$ is fixed and not scaled with the amount of initially deposited antigen 
($\Theta_{\rm ag}=N_{\rm ag}$), then such robustness can be observed on a low level (see Figures 5 and 6 in the Supplementary Information). It should be noted that successful affinity maturation at lower antigen densities results from a reduced binding probability for all affinities (as discussed in the previous section) such that the observed robustness is the result of two different mechanisms acting in different antigen density ranges.

We conclude that {\it in silico} antigen consumption does not improve affinity maturation over site competition alone. At low antigen densities the general properties of the GC reaction can no longer be reproduced. Affinity maturation is therefore unlikely to be driven by antigen consumption even when combined with site competition.

\subsection{The impact of antigen masking by competing antibodies}
Plasma cells outside the GC produce antibodies at rates of up to 
$r_{\rm ab}=2\cdot 10^3$ antibody molecules per second \cite{randall92}. These antibodies may mask antigen in the GCs and thereby increase the selection pressure on B cells and simultaneously provide a feedback mechanism that enables robustness towards alterations in the initial antigen density \cite{vora97,tarlinton00,iber02} (see also Figure 1D in the Supplementary Information). Assuming that plasma cells produce high affinity antibodies in the GC at a rate of $r_{\rm ab}=10^2$ per output cell (thus overestimating a realistic production of soluble antibody in GCs), we find that the above considerations do indeed hold true: A physiological GC kinetic can be obtained for antigen amounts between 20 and 200 antigen portions per FDC site (Figure 7 in the Supplementary Information). A small improvement in affinity maturation is observed relative to site and antigen competition alone (Figure 5). As before, affinity maturation approaches the site competition value for 120 sites when antigen is abundant (Figure 5B and 1B).
\begin{figure} [!ht]
\begin{minipage}{15cm}
\begin{center}
\includegraphics[width=7.0cm]{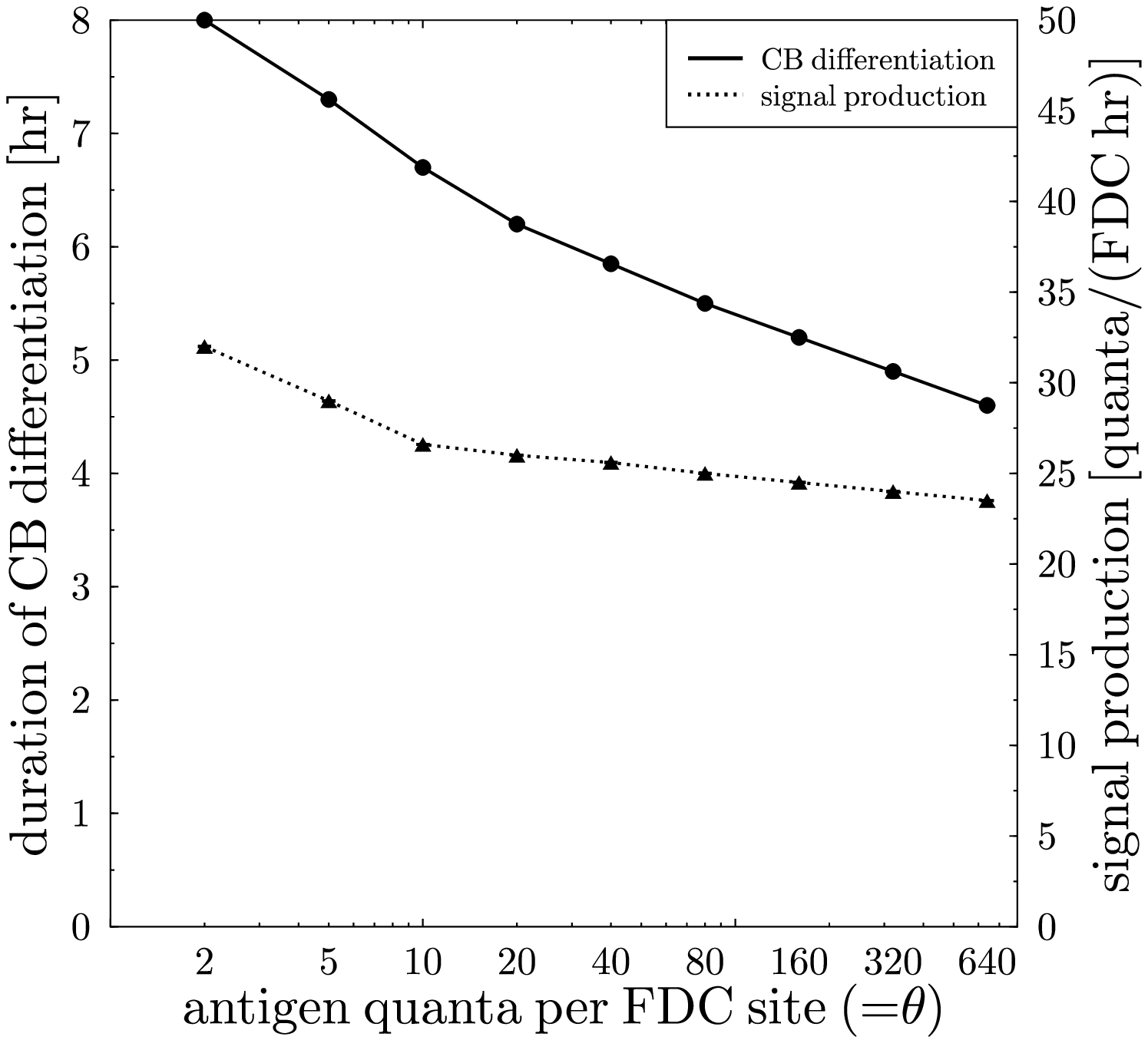}
\includegraphics[width=7.0cm]{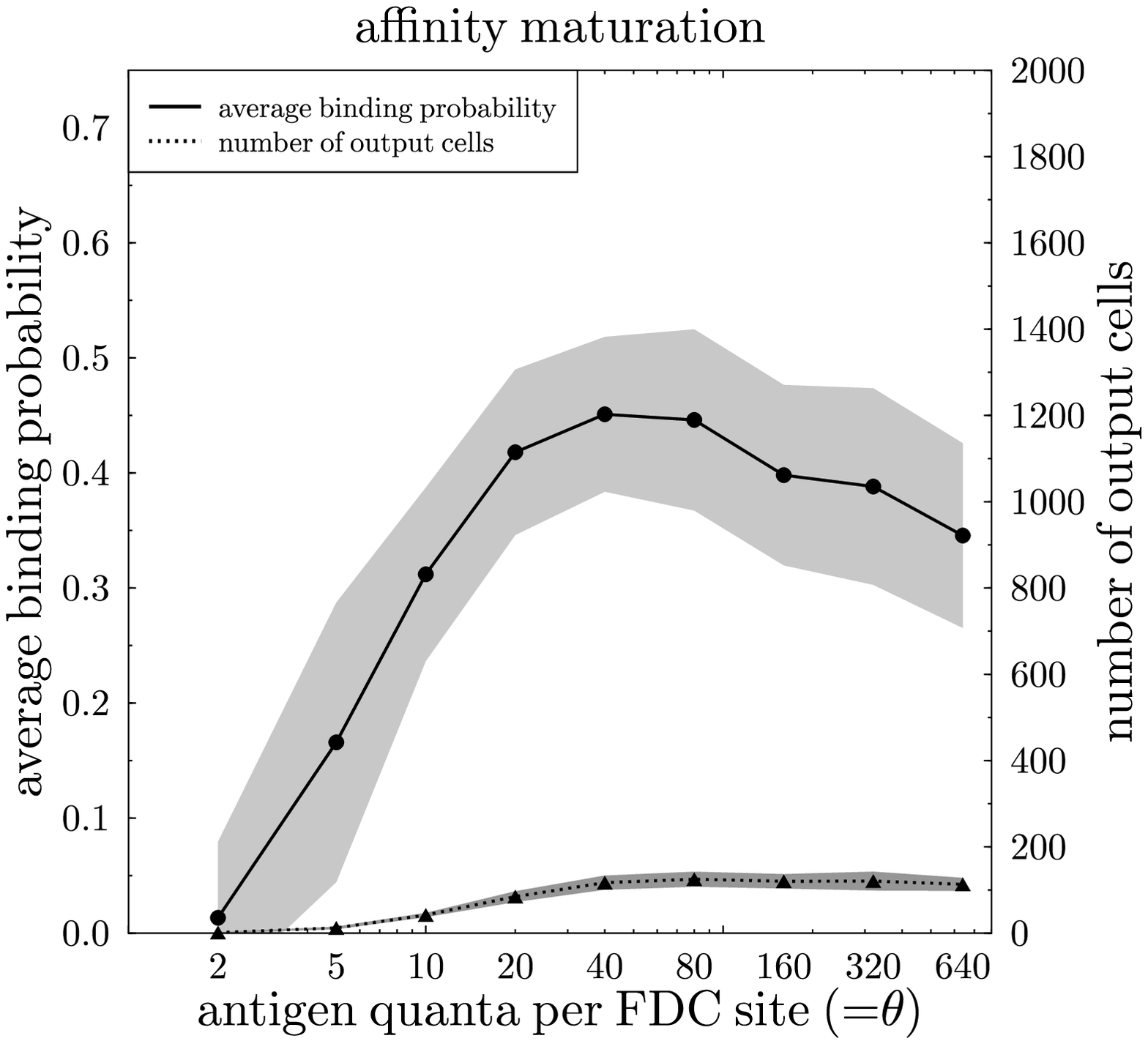}\\[-6.5cm]
\hspace*{-6.5cm}{\bf A}
\hspace*{7.0cm} {\bf B}\\[6.0cm]
\caption[]{\label{Fig_abtx} \baselineskip12pt {\small 
\textbf{Antigen masking by
soluble antibodies increases competition for antigen in the early
phase of GC reaction.} 
{\it In silico} GC experiments as in Figure 4 with antibody production of $10^2$ antibodies per second and per plasma cell (see Figure 7 in the Supplementary Information for the full dataset). {\bf (A)} Parameter adaptation: {\it duration of CB differentiation} and differentiation {\it signal production} rate per FDC, {\bf (B)} extent of affinity maturation (measured as average binding probability and number of output cells).}}
\end{center}
\end{minipage}
\end{figure}

The effect of strong antibody production is an early reduction in free antigen, and thus an early increase in the selection pressure (Figure 3C). For example, in the case of 40 antigen portions per FDC site (i.e.~for optimal affinity maturation) only 25\% of the antigen is used during the GC reaction, but free antigen is kept in the limiting regime throughout the reaction (Figure 3C). In that way the inefficient time course of affinity maturation found in the case of antigen consumption without soluble antibodies (Figure 3B) is restored to some extend.

While antigen masking leads to the correct kinetics and a small improvement of affinity maturation, the number of produced output cells is rather small (Figure 5B), as is the ratio of CC to CB, and $q^{12}_6\ll 6$ (Figure 5B in the Supplementary Information). We conclude that antigen masking resolves some of the problems in the antigen competition scenario, but some major GC properties are still not reproduced. In addition this mechanism is not robust to a substantial (10-fold) decrease in the antibody production rate, which is more realistic than the rate assumed in the present {\it in silico} experiment.

\subsection{Competition for T cell help}
Interactions between antigen specific CCs and T cells are necessary for CC survival and differentiation \cite{devinuesa00}. Given that T cells only constitute 5-10\% of the GC cell population \cite{kelsoe96} this interaction may be competitive. While T cells can bind to several B cells simultaneously, they can only polarize to one \cite{kupfer94}, which we assume to be the CC that has previously bound antigen with highest affinity (see also Figure 1E in the Supplementary Information). In our simulation, polarization has to be maintained for a minimum time 
($\Delta t_{\rm rescue}$) to rescue a B cell. Cells that bind to T cells for longer than $\Delta t_{\rm apop} > \Delta t_{\rm rescue}$ and do not receive rescue signals for at least $\Delta t_{\rm rescue}$ are doomed to apoptosis. Note that all CCs need to be selected by FDCs before they can compete for T cell help.

We find that the extent of affinity maturation is only slightly enhanced as compared to site competition alone ($X_{\rm FDC}=120$ sites) if 5-10\% of all GC cells are antigen-specific T cells (50-100 T cells per GC in the two-dimensional simulation) to which antigen-presenting CCs can bind (Figure 6B full line). However, given that each antigen-specific T cell only recognizes a certain antigen epitope and different CCs will process the antigen differently and therefore present different epitopes, only a small fraction of GC T cells will be reactive to a given CC. The simulation reveals that affinity maturation is strongest (60-70\% of all output cells are of high affinity) when only about 0.5-1\% of all GC cells (5-10 T cells per GC in the two-dimensional simulation) can rescue a given antigen presenting CC (Figure 6B full line). The GC characteristics are reproduced (Figure 6A, and Figure 8 in the Supplementary Information). The mechanism is robust to large variations in antigen availability and may still work when antigen is encountered in soluble form since the simulations were performed in a regime where antigen acquisition and interactions with FDCs are uncompetitive.
\begin{figure} [!ht]
\begin{minipage}{15cm}
\begin{center}
\includegraphics[width=7.0cm]{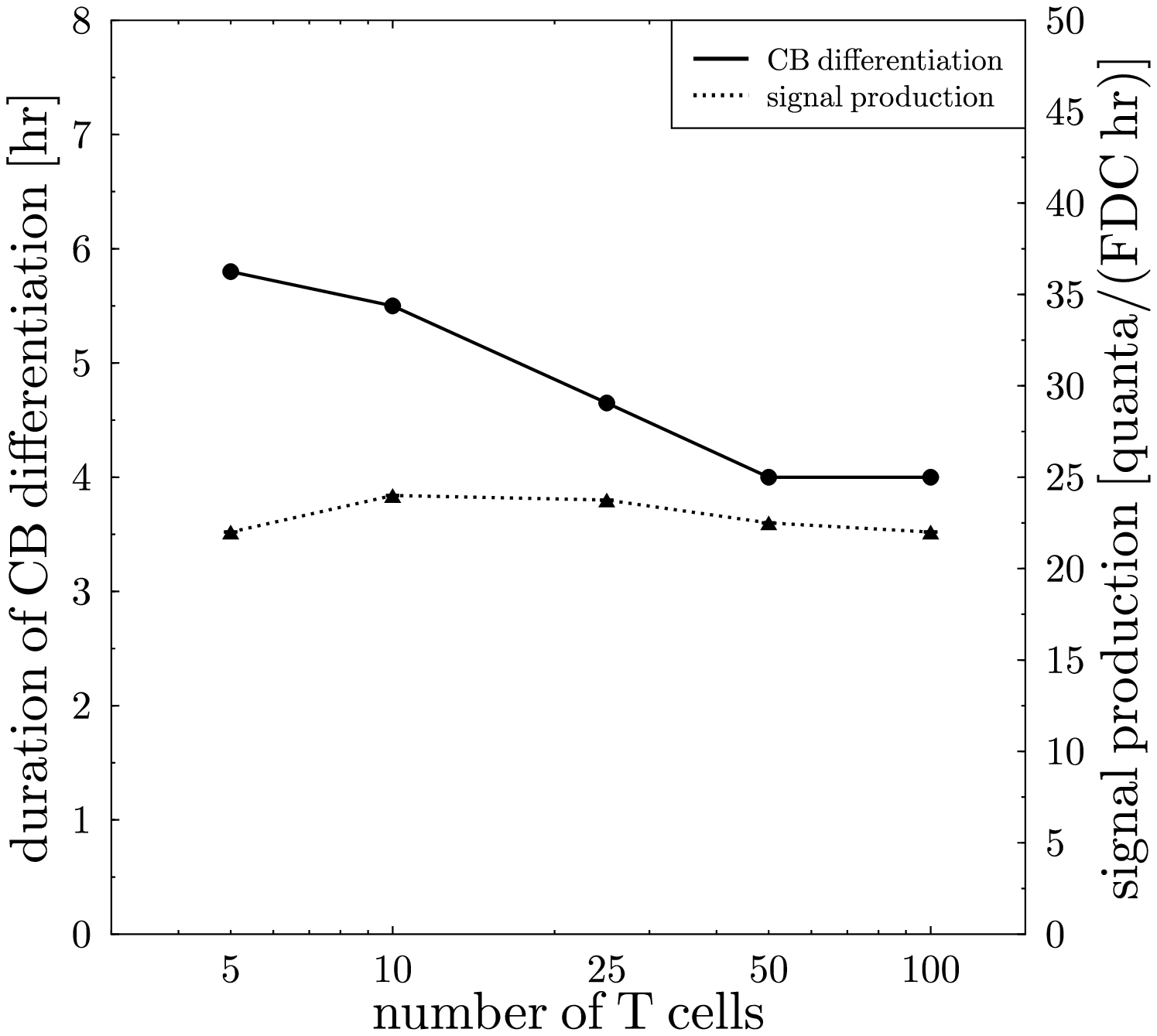}
\includegraphics[width=7.0cm]{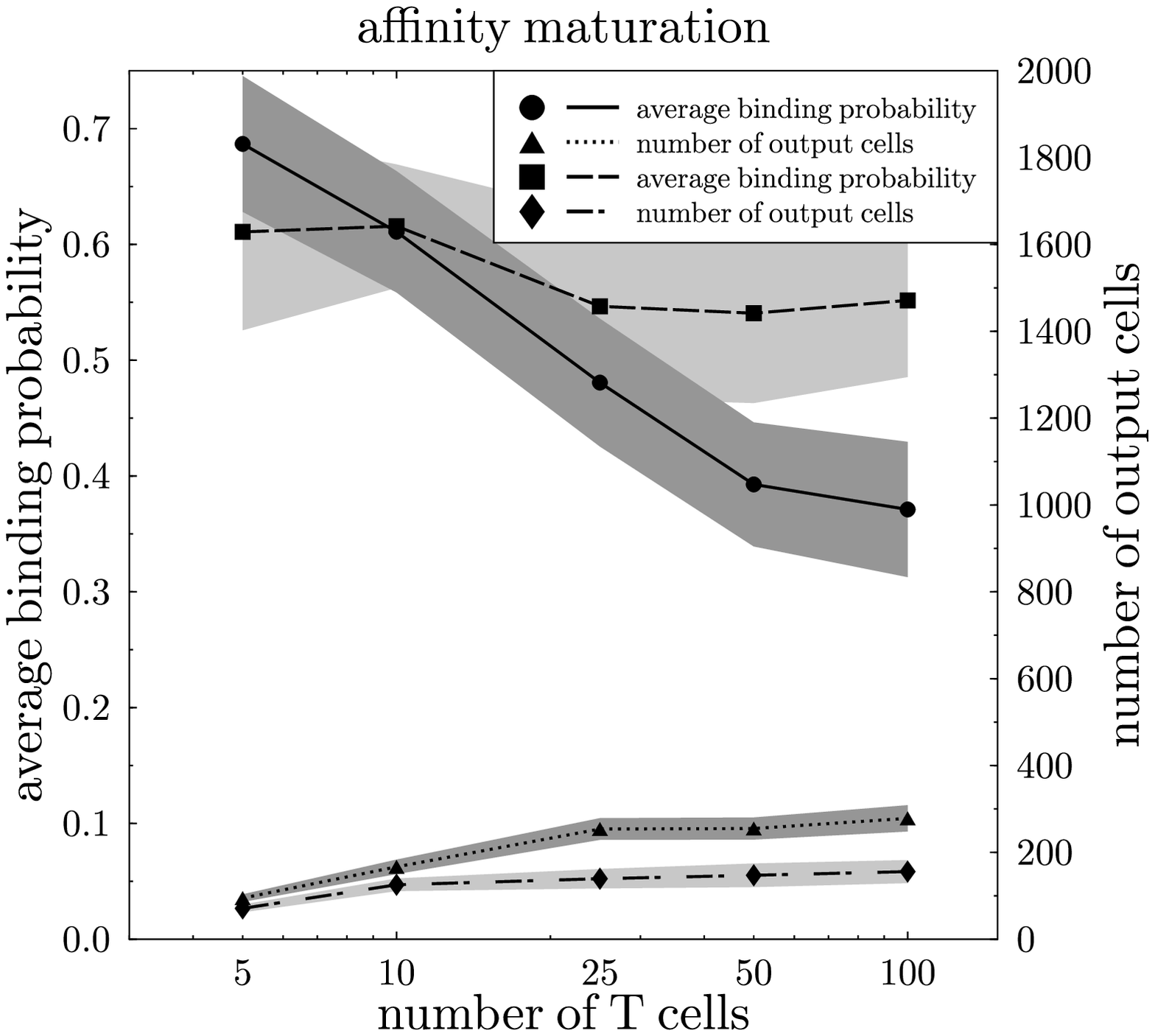}\\[-6.5cm]
\hspace*{-6.5cm}{\bf A}
\hspace*{7.0cm} {\bf B}\\[6.0cm]
\caption[]{\label{Fig_d_tc} \baselineskip12pt {\small 
\textbf{Competition for T cell help allows for robust aff\/inity maturation.} 
{\it In silico} GC experiments with 120 FDC sites, abundant antigen, and T cells as indicated on the axis 
($\Delta t_{\rm rescue}=2$ hours and $\Delta t_{\rm apop}=2.1$ hours) (see Figure 8 in the Supplementary Information for the full dataset). {\bf (A)} Parameter adaptation: {\it duration of CB differentiation} and differentiation {\it signal production} rate per FDC, {\bf (B)} extent of affinity maturation (measured as average binding probability and number of output cells. Combining competition for T cell help with a CC refractory time of 0.5 hours and a reduction of the binding probability to 0.5 leads to more robust affinity maturation (squares and diamonds, and Figure 11 in the Supplementary Information) than in the standard case (circles and triangles).}}
\end{center}
\end{minipage}
\end{figure}

The time span $\Delta t_{\rm apop}$ for which CCs can survive while binding to T cells without receiving rescue signals has a major impact on the level of affinity maturation. Thus a small increase in $\Delta t_{\rm apop}$ from 2.1 to 4 hours already reduces affinity maturation to a level comparable to site competition (see Figure 9 in the Supplementary Information). A combination of T cell competition with a general reduction of the binding probability to 50\% of its value (see above) has no major effect on the result (see Figure 10 in Supplementary Information). However, a combination of competition for T cell help with a longer CC refractory time makes affinity maturation more robust against changes in the T cell density (Figure 6B dashed line).

\section{Discussion}
The simulation results presented here reveal current models of B cell selection (competition for access to FDCs or antigen) to be incompatible with available experimental information. We propose two novel mechanisms, a refractory time between CC-FDC engagements as well as competition for T cell help. Both mechanisms enable strong affinity maturation {\it in silico} while reproducing all available experimental information on GC dynamics. In agreement with experiments \cite{vora97,hannum00}, the mechanisms are robust to variations in the amount of initially deposited antigen. As competition for antigen held on FDC is assumed to be non-limiting in both mechanisms, it is likely that these mechanisms also work with soluble antigen which would be in agreement with results from recent experiments that call into question the concept that antigen presentation by FDCs is essential \cite{hannum00,haberman03}. However, in the present simulations an FDC-independent pre-selection of CCs with soluble antigen has not been tested. Note that this {\it in silico} simulation is not able to cover all possible cellular interactions that might serve as suitable mechanism for affinity maturation. Thus the two highlighted novel mechanisms have to be considered as propositions.

Competition for access to FDCs or antigen does not provide a sufficient selection pressure to enable a high level of affinity maturation in the simulation. While antigen consumption does lead to an increase in the selection pressure, this increase comes too late so that many low affinity output cells have already been produced and the remaining cell numbers are insufficient for successful affinity maturation. Moreover, for low antigen densities the GC reaction ends prematurely since antigen consumption then removes too much antigen. While antigen masking by emerging antibodies can, in principle, increase the selection pressure earlier, the antibody production rate would need to be unphysiologically high. Even then the available data on GCs could not be reproduced and, in addition, affinity maturation is not as strong as that produced by the alternative mechanism. Given that evolution tends to optimise performance, this latter limitation may also be interpreted as an argument against these mechanisms.

At first sight, less than 200 output cells (that is less than 8000 cells in three dimensions) may seem low. Given the efficient expansion of output cells the success of an immune response can, however, be expected to primarily depend on the size of the high affinity fraction; this is therefore a physiologically realistic result. Note that the level of affinity maturation that is achieved in the simulations (60-70\% of all output cells are of high affinity) is in qualitative agreement with the experimentally observed domination of high affinity clones \cite{smith97}. We had to restrict ourselves to two-dimensional simulations to generate the results with sufficient statistics, which prevents us from a quantitative comparison of this result.

To our knowledge this theoretical investigation is the first systematic comparison of different B cell selection mechanisms in GC reactions as well as the first theoretical work to investigate a possible role of T cells in the selection process. Previous models have concentrated on single selection mechanisms \cite{iber02,meyer-hermann02_jtb,meyer-hermann05,oprea97,oprea00,beyer02,kesmir99,meyer-hermann01} and have, in general, described them only phenomenologically, i.e.~without including details of cell interactions. The impact of the spatial cell distribution has only been investigated in a few models \cite{kesmir03,meyer-hermann02_jtb,meyer-hermann05,beyer02}. The present model is based on available GC kinetics and cell motility data (see Model section) and the parameters have been evaluated carefully using data from mice and rat lymph nodes (see Table 1). We did not find any experimentally uncertain parameter which would change our results qualitatively.

The only critical assumption is the shape space concept combined with a smooth affinity function. Changes of the width of the function for the affinity dependent selection probability in the shape space has important consequences for GC development. For small variations of the width all constraints from experiment can be respected by a corresponding adaption of other parameters. However, this becomes impossible when the width is strongly varied. This shows that our choice for the width is well-defined. While the suitability of the shape space concept for antigen-antibody affinity may be questioned, insufficient molecular information on antigen-antibody interaction is currently available to allow a more accurate model of mutation-dependent increases in antibody-antigen affinity.

Autoreactive BC have not been considered in the present simulations because under normal circumstances they are not positively selected by T cells and are therefore eliminated by apoptosis similar to other antigen-unspecific CCs. We therefore assumed autoreactive BC to be part of the pool of low affinity BC that emerge from the seeder cells by disadvantageous mutations. This may shift the proportion of advantageous and disadvantageous mutations. However, we believe that this approximation remains within the error introduced by the shape space concept itself which most likely will inaccurately reproduce this proportion. 

The CC refractory time and the competition for T cell help are equally attractive selection mechanisms to explain antibody affinity maturation in GCs and are both likely to play a role in B cell selection. The CC refractory time limits the number of selection trials per CC lifespan such that only high affinity B cells which bind with high probability on the first or second trial are efficiently selected. The CC refractory time will be a consequence of IgM-independent interactions between CCs and FDCs. While the exact duration of such interactions still has to be determined in studies that employ CCs and FDCs, a refractory time between two CC-FDC encounters as long as 4 hours appears to be unlikely. However, we have also shown that an additional general (i.e.~affinity independent) reduction of the CC-FDC binding probability to 0.5 may reduce the necessary refractory time to 15-30 minutes. One may speculate that integrin interactions may result in such a refractory time. LFA-1/ICAM-1 and VLA-4/VCAM-1 have been described to mediate adhesion of B cells to FDCs \cite{koopman91}. Even though unspecific contacts are dissolved in the range of minutes \cite{gunzer04}, experiments with lipid bilayers have also shown that especially at low to medium affinity antigen binding may fail to induce the formation of an immunological synapse but can still result in a firm integrin-dependent attachment of B cells \cite{carrasco04}. The speed at which these integrin-dependent contacts form and dissolve if B cells fail to establish an immunological synapse (which is generally believed to enable B cell activation and antigen extraction \cite{batista01}) may determine the refractory time. A different possibility is a reduced CC motility after an unsuccessful binding event. This might reduce the subsequent interaction frequency of CCs with FDCs.\footnote{We thank the reviewer for this suggestion}. 

An additional competition for T cell help further reduces the necessary CC refractory time. It has been shown {\it in vitro} that specific T cell help is essential for affinity maturation in the case of NP ((4-hydroxy-3-nitrophenyl)acetyl) \cite{aydar05}, and more generally the role of T cells in the selection process has been highlighted \cite{devinuesa00}. While there is as yet neither direct experimental evidence for competition for T cell help nor have antigen-specific T helper cells in the GC so far been suggested as a limiting factor for affinity maturation, what is known about B cell-T cell interactions strongly supports such a model. Thus T cells have been observed to interact with several B cells at a time but to only activate the B cell towards which the microtubule organizing center (MTOC) (and the colocalized Golgi apparatus) are reorganized \cite{kupfer94,poo88,kupfer91}. 
Reorientation of the MTOC is dependent on and directed towards the site of T
cell receptor signaling \cite{sedwick91}, and, when interacting with
several B cells, T cells polarize to the B cell with the highest density of
specific pMHC \cite{depoil05}. We propose that a higher affinity of
interaction results in increased pMHC presentation such that only the
highest affinity-CC interacting with a T cell is rescued. Such causal
relationship is in agreement with the earlier observations that T cell
responsiveness to B cells depends on both the antigen concentration and
affinity \cite{batista98} as well as on the (antigen availability
dependent) number of MHC class II loaded with antigenic peptide(pMHC) that
are presented on the B cell surface \cite{harding90,grakoui99}. 
It is not an essential element of the present model but a plausible
hypothesis that apoptosis is induced in all other B cells to which the T
cell does not sufficiently polarise.
This negative selection by T helper cells could in principle be mediated by FasL-Fas as interactions which can be expected to form in all contact zones independent of T cell polarization. Whether FasL is at all upregulated on GC T cells is, however, currently unclear since the spatial distribution of FasL in GCs from patients with lymphofollicular hyperplasia correlates with the FDC network rather than with the distribution of CD4 positive T cells \cite{verbeke99}.

While simultaneous interactions of T cells with several B cells in a lymph node environment have been reported as transient \cite{okada05}, all B cells and T cells used in that study were of the same specificity and T cells rather than B cells were abundant. It will be important to repeat these studies under conditions that are more representative of GC reactions and to access the affinity dependence of T cell-B cell interactions in far greater detail. Imaging of T cell interactions with B cells that have recognized antigen with different affinity should help to clarify whether the MTOC does indeed reorganize to the highest affinity B cell and whether Fas-FasL accumulate in the contact zones to induce apoptosis in the outcompeted CC. Moreover, it will be interesting to test whether an (artificial) expansion of antigen-specific GC T cell clones can hamper antibody affinity maturation.

B cell selection in competition for T cell help challenges the presently held view that FDCs play a pivotal role in B cell selection. In this novel mechanistic framework FDCs only play a role in supporting fragile CCs with general survival and differentiation signals but do not drive affinity maturation by negatively selecting B cells.

A competition for T cell help and a long CC refractory time both enable affinity maturation under conditions when acquisition of FDC signals is uncompetitive such as for a wide range of antigen concentrations. Accordingly, these mechanisms are in agreement with experiments that show that a reduction of antigen densities on FDCs by at least 400-fold does not prevent antibody affinity maturation \cite{hannum00}. 

We conclude that within the selection mechanisms tested, thus using a restricted set of possible interaction mechanisms, a combination of T cell competition together with slow CC-FDC interactions and a sub-optimal binding probability of high affinity clones can drive B cell selection in GCs under the experimentally established conditions. Further insight into GC dynamics and affinity maturation should be gained by combining the powerful predictions of these agent based simulations with the latest imaging techniques.

\section*{Acknowledgement}
M.M.-H.~was supported by a Marie Curie Intra-European Fellowship 
within the Sixth EU Framework Program.
D.I. is a Junior Research Fellow at St John's College, University of Oxford
and is supported by a \mbox{DTA} \mbox{EPSRC} studentship.
FIAS is supported by the ALTANA AG.

\bibliographystyle{jtb}
\bibliography{refs_MMH}

\end{document}